\pdfoutput=1%
\documentclass[a4paper,aip,american,citeautoscript,floatfix,jcp,longbibliography,pdftex,reprint,superscriptaddress,twocolumn]{revtex4-1}
\usepackage{amsfonts,amsmath,amssymb}%
\usepackage{graphicx}%
\usepackage[utf8]{inputenc}%
\usepackage[T1]{fontenc}
\usepackage{textcomp}%
\usepackage[textsize=footnotesize]{todonotes}%
\usepackage{xspace}
\usepackage[breaklinks=true]{hyperref}%
\usepackage{breakurl}%
\usepackage{hypernat}%
\usepackage{soul}%

\newlength{\figwidth}%
\newlength{\figwidthlarge}%
\newlength{\figwidthsmall}%
\newcommand{\bohr}{\ensuremath{\text{a}_0}\xspace}%
\newcommand{\degree}{\ensuremath{^\circ}\xspace}%
\newcommand{\eg}{e.\,g.}%
\newcommand{\etal}{et al.}%
\newcommand{\ie}{i.\,e.}%
\newcommand{\mueff}{\ensuremath{\mu_{\text{eff}}}}%
\newcommand{\qcms}{\ensuremath{\textup{cm}^3\,\textup{s}^{-1}}}
\newcommand{\Trot}{\ensuremath{T_\textup{rot}}\xspace}%

\newcommand{\AP}{\ensuremath{\text{3AP}}\xspace}%
\newcommand{\CaForty}{\ensuremath{{}^{40}\text{Ca}^+}\xspace}%
\newcommand{\Ca}{\ensuremath{\text{Ca}^+}\xspace}%
\newcommand{\cis}{\textit{cis}\xspace}%
\newcommand{\NNO}{\ensuremath{\text{N}_2\mathrm{O}}\xspace}%
\newcommand{\trans}{\textit{trans}\xspace}%

\newcommand{\cfel}{\affiliation{Center for Free-Electron Laser Science, DESY, Notkestrasse 85, 22607 Hamburg, Germany}}%
\newcommand{\cui}{\affiliation{The Hamburg Center for Ultrafast Imaging, University of Hamburg, Luruper Chaussee 149, 22761 Hamburg, Germany}}%
\newcommand{\ubas}{\affiliation{Department of Chemistry, University of Basel, Klingelbergstrasse 80, 4056 Basel, Switzerland}}%
\newcommand{\uhh}{\affiliation{Department of Physics, University of Hamburg, Luruper Chaussee 149, 22761 Hamburg, Germany}}%

\begin{document}
\title{ Chemical reactions of conformationally selected molecules in a beam with Coulomb-crystallized ions}%
\author{Daniel Rösch}\ubas%
\author{Stefan Willitsch}\email{stefan.willitsch@unibas.ch}\ubas%
\author{Yuan-Pin Chang}\cfel%
\author{Jochen Küpper}\email{jochen.kuepper@cfel.de}\cfel\uhh\cui%
\date{\today}%
\begin{abstract}\noindent%
   Many molecules exhibit multiple conformers that often easily interconvert under thermal
   conditions. Therefore, single conformations are difficult to isolate which renders the study of
   their distinct chemical reactivities challenging. We have recently reported a new experimental
   method for the characterization of conformer-specific effects in chemical reactions [Y. P. Chang
   et al., Science 342, 98 (2013)]. Different conformers are spatially separated using inhomogeneous
   electric fields and reacted with a Coulomb crystal of cold, spatially localized ions in a trap.
   As a first application, we studied reactions between the two conformers of 3-aminophenol and \Ca.
   We observed a twofold larger rate constant for the \cis compared to the \trans conformer which
   was rationalized in terms of the differences in the long-range ion-molecule interactions. The
   present article provides a detailed description of the new method and a full account of the
   experimental results as well as the accompanying theoretical calculations.
\end{abstract}
\maketitle%

\section{Introduction}
\label{sec:intro}

Many molecules possess multiple conformers (rotational structural isomers) that interconvert with
low energy barriers through hindered rotations about single covalent bonds. It is well known that
different conformations can exhibit distinct chemical reactivities~\cite{Barton:JCS1027,
   Dunathan:PNAS55:712, Eliel:StereoChem:1994}. To gain a comprehensive understanding of the
chemical behavior of molecules, it is thus necessary to investigate their conformer-specific
chemistry. To this end, individual molecular conformations need to be isolated and characterized.
Moreover, the capability to manipulate conformational distributions provides means to influence
reactivities and products, adding to the repertoire of methods to control chemical processes.

When studies of conformational effects are carried out at low temperatures in the gas phase, the
thermal interconversion of conformations is suppressed. In recent years, significant progress toward
the spectroscopic characterization of specific conformations has been
achieved~\cite{Suenram:JACS102:7180, Rizzo:JCP84:2534, Robertson:PCCP3:1, Weinkauf:EPJD20:309,
   Simons:PCCP6:Biomolecules, Simons:IRPC24:489, Vries:ARPC58:585, EugeniaSanz:ACIE47:6216,
   Rizzo:IRPC28:481, Nagornova:Science336:320}. For instance, studies of photoinduced ground-state
isomerization reactions mapped out interconversion pathways between conformations, \ie, minima on
the corrugated potential energy surfaces~\cite{Dian:Science296:2369, Dian:Science303:1169,
   Dian:Science320:924}. Investigations of the conformational landscape of small molecules, \eg,
poly-aminoacids, helped to understand the folding motives in peptides~\cite{Vries:ARPC58:585}.

Unimolecular conformer-specific dynamics have been investigated in the photodissociation of cations
of small organic molecules~\cite{Park:JACS124:7614, Park:Nature415:306, Kim:Science315:1561}. In
these studies, individual conformers of cations were selectively generated capitalizing on the
different ionization energies of their parent neutrals. In addition, photodissociation studies on
neutral conformers of small organic molecules have been carried out~\cite{Oliver:JCP133:194303,
   Oliver:CS1:89, Zaouris:JCP135:094312}. In these studies, individual conformers were not separated
prior to photoexcitation. Consequently, the experimental results contained contributions from all
populated conformations. These were disentangled through Rydberg tagging methods which allowed for
resolving the small energy difference between conformers manifested in the kinetic energy release of
H photofragments.

For bimolecular reactions in the gas phase only few investigations of conformational effects have
been reported so far. Taatjes \etal~\cite{Taatjes:Science340:177} observed conformer-dependent
reactivities from the simplest Criegee intermediate (CH$_3$OO) which plays a key role in ozonolysis
reactions. In a cryogenic matrix, Khriachtchev \etal~\cite{Khriachtchev:JPCA113:8143} observed
distinct conformer-dependent products from two formic acid conformers reacting with oxygen atoms.

Several techniques for the manipulation of conformational distributions of molecules in the
gas-phase have been reported. Ion mobility allows the separation of molecules according to their
shape and thus serves as a key technique to separate classes of conformers of large charged
molecules~\cite{Helden:Science267:1483}. For neutral molecules, the spatial separation of specific
conformers has been achieved using electric fields~\cite{Filsinger:PRL100:133003,
   Filsinger:ACIE48:6900, Kierspel:CPL591:130}. While all conformers of a molecule have the same
mass, they often differ by their dipole moments. Different dipole moments lead to different Stark
shifts of rotational energy levels in electric fields. Thus, in an inhomogeneous electric field
different forces act on the different conformers, which can be used for the manipulation of their
translational motion.

Recently, we have adapted this technique to investigate the chemical reactivities, \ie, rate
constants, of specific conformers in the prototypical bimolecular reactions of 3-aminophenol (\AP)
and \Ca~\cite{Chang:Science342:98}. \AP has two stable conformers (denoted \cis and \trans) which
differ in the orientation of the OH group and have significantly different dipole moments (2.33~D
and 0.77~D for the \cis- and \trans-species, respectively). \AP was entrained in a molecular beam
and spatially separated using an electrostatic deflector. The dispersed molecular beam was directed
at a stationary reaction target consisting of a Coulomb crystal of \Ca ions, \ie, an ordered
structure of translationally cold ions at a temperature of a few millikelvins in a
trap~\cite{Willitsch:IRPC31:175}. Singly ionized Ca ions were chosen as a co-reactant because they
can easily be Coulomb-crystallized by laser cooling. \Ca is also known for its reactivity with
organic molecules acting as a catalyst for the activation of inert chemical
bonds~\cite{Eller:CR91:1121, Schwarz:ACIE50:10096} such as C-F and C-O~\cite{Harvey:CPL273:164,
   Zhao:JPCA110:10607, Ryzhov:JACS121:2259}.

In the present article, we give a detailed account of our methods and results on the
conformer-specific reactivities of \AP with \Ca. The outline of the paper is as follows:
\autoref{sec:setup} and \autoref{sec:theory:method} describe the experimental setup and theoretical
procedures employed. In \autoref{sec:exp_result}, we present a characterization of the electrostatic
deflection of the conformers, their reaction profiles, conformer-specific reaction rate constants as
well as mass spectra of the reaction products. An analysis of the results based on theoretical
calculations follows in \autoref{sec:theory:results}.

\section{Experimental Setup}
\label{sec:setup}
\begin{figure*}
   \includegraphics[width=\textwidth]{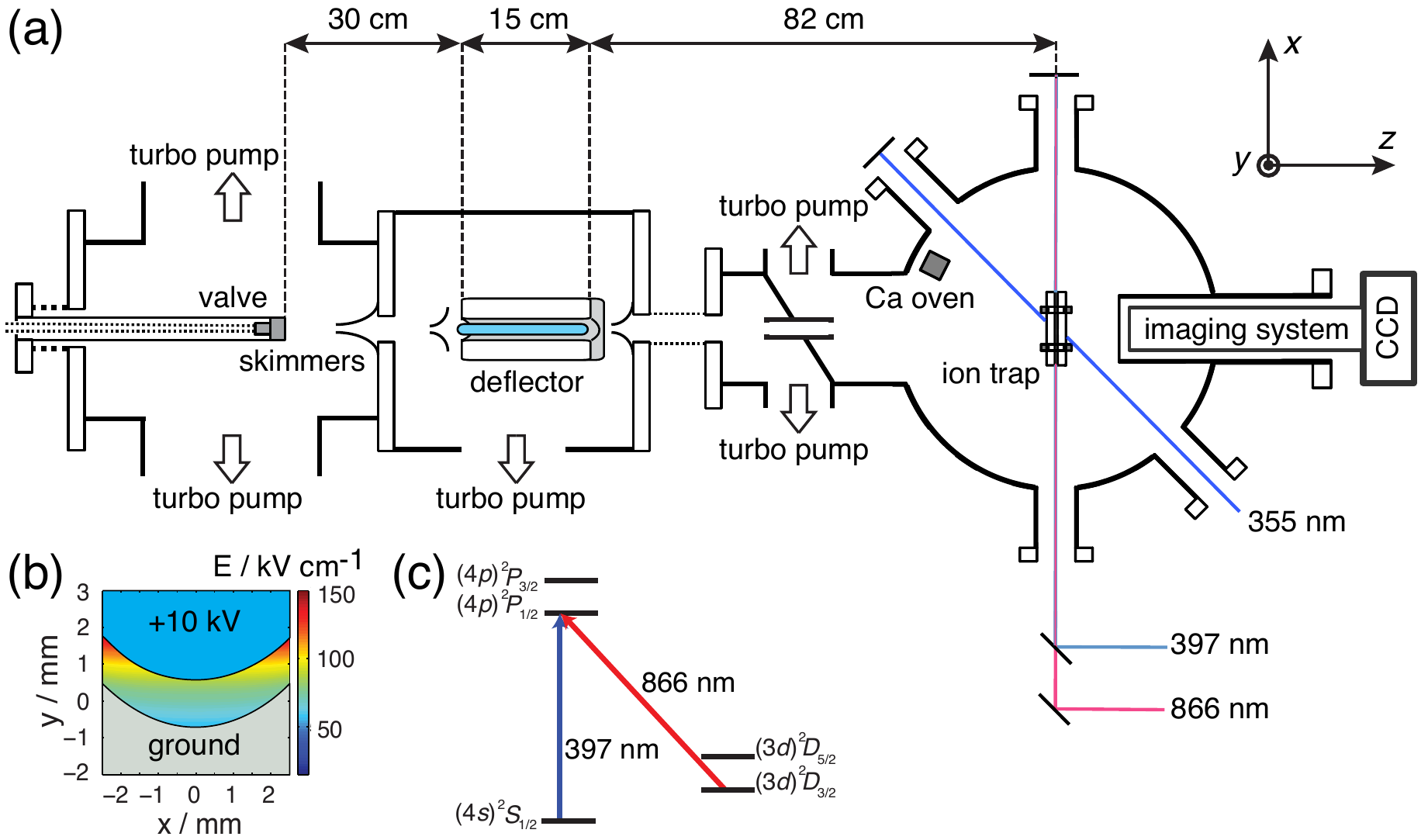}
   \caption{(a) Schematic top view of the experimental setup for studying conformer-selected
      chemical reactions. See text for details. (b) Electric field strength $E$ along a cut through
      the electrostatic deflector. (c) Diagram of energy levels accessed during Doppler laser
      cooling of \CaForty.}
   \label{fig:setup}
\end{figure*}
The experimental setup consists of two main parts: a molecular beam deflection apparatus for the
separation of \AP conformers and an ion trap apparatus for the generation and storage of Coulomb
crystals of laser cooled \Ca ions. The individual experimental procedures have been reported
previously~\cite{Filsinger:JCP131:064309, Filsinger:ACIE48:6900, Willitsch:PCCP10:7200,
   Willitsch:IRPC31:175, Chang:Science342:98}. In the following, we focus on the details of the
combined apparatus and the methodology for conformer-specific reaction experiments.

\subsection{Conformer deflection setup}
\label{sec:setup:molbeam}
The molecular beam machine for conformer deflection consisted of a series of differentially pumped
vacuum chambers. The source chamber housing a pulsed valve was pumped by two 1650~l/s turbomolecular
pumps. The deflector chamber containing the electrostatic deflector was pumped by a 500~l/s
turbomolecular pump, as shown in \autoref{fig:setup}\,(a). A solid sample of \AP (Sigma-Aldrich,
$98~\%$) was placed in a reservoir cartridge and vaporized at $145$~\degree{C} inside a
high-temperature Even-Lavie valve~\cite{Even:JCP112:8068}. The valve was operated at a backing
pressure of $35~$bar of neon at a repetition rate of 600~Hz. The typical rotational temperature of
\AP in our experiments was about $1$~K. Two skimmers with diameters of $2~$mm and $1$~mm were placed
$15$~cm and $27$~cm downstream from the nozzle, respectively. After skimming, the collimated
molecular beam entered the 15~cm long electrostatic deflector~\cite{Hughes:PR72:614, Lee:PR91:1395,
   Ramsey:MolBeam:1956, Holmegaard:PRL102:023001}. A cut through the electrodes of the deflector
including a contour plot of the generated electric field is shown in \autoref{fig:setup}. The
vertical gap between the deflector electrodes perpendicular to the molecular beam axis was $1.4~$mm.
The shape of the electrodes was designed to generate a strong inhomogeneous electric field with a
nearly constant gradient along the $y$ axis~\cite{Hughes:PR72:614, Lee:PR91:1395}. The molecular
beam passed a third skimmer with a diameter of 1.5~mm for differential pumping into a chamber pumped
by a 345~l/s turbomolecular pump. Subsequently, the beam entered the reaction chamber, pumped by a
550~l/s turbomolecular pump, through another differential pumping aperture formed by a 3.5~cm long,
10-mm-diameter tube. Typical pressures during the experiments were $7 \cdot 10^{-6}$~mbar, $9 \cdot
10^{-8}$~mbar, and $2 \cdot 10^{-9}$~mbar in the source, deflector and reaction chambers.

\subsection{Ion trap setup}
\label{sec:setup:iontrap}
In the reaction chamber, Coulomb crystals of laser-cooled \Ca ions were generated and trapped in a
linear radiofrequency (RF) ion trap~\cite{Willitsch:PCCP10:7200, Willitsch:IRPC31:175}. \Ca ions
were produced by non-resonant multi-photon ionization of a beam of Ca atoms evaporated from an oven
and passing through the center of the ion trap~\cite{Hall:PRL107:2432002, Willitsch:IRPC31:175}, see
\autoref{fig:setup}\,(a). The ion trap consisted of four segmented cylindrical electrodes with a
radius $r=4.0$~mm arranged in a quadrupolar configuration. To confine the ions in the plane
perpendicular to the trap symmetry axis, RF voltages with amplitudes $V_{\text{0,RF}}=350$~V and
frequencies $\Omega=2\pi\times3.1$~MHz were applied with opposite polarities across adjacent
electrodes. To confine the ions along the axis, static voltages in the range of
$V_{\text{end}}=1$--10~V were applied to the endcap electrodes. The atomic beam was ionized using
the third harmonic (355~nm) of a Nd:YAG laser close to the center of the ion trap. The \Ca ions were
laser-cooled with beams produced by two external cavity enhanced diode lasers operating at
wavelengths of 397~nm and 866~nm to pump the $(4s)~^2S_{1/2}\rightarrow(4p)~^2P_{1/2}$ and
$(3d)~^2D_{3/2}\rightarrow(4p)~^2P_{1/2}$ transitions, respectively~\cite{Willitsch:IRPC31:175}, see
\autoref{fig:setup}\,(c). The frequencies of the two laser beams were simultaneously monitored using
an automated fiber-switcher coupled to a wavemeter and stabilized by a computer-controlled voltage
feedback loop. The resulting laser linewidths were on the order of a few MHz. The laser powers
employed were about $600~\mu$W and $200~\mu$W for the 397~nm and 866~nm beams, respectively. Upon
laser cooling, the ions localized in space and formed three dimensional spheroidal Coulomb
crystals~\cite{Willitsch:PCCP10:7200, Willitsch:IRPC31:175} with a radius $r\approx200$~$\mu$m and a
width $z\approx550$~$\mu$m typically consisting of $\sim\!700$ ions. The secular kinetic energy of
the laser cooled ions amounted to $E_{\text{sec}}\approx{}k_\text{B}\cdot10$~mK. Two-dimensional
cuts of the central plane of the Coulomb crystals were imaged by collecting a solid angle of the
atomic fluorescence generated during laser cooling using an enhanced CCD camera coupled to a
microscope with ten-fold magnification.

\subsection{Reaction rate measurements}
\label{exp:rate}
The first step in each reaction experiment consisted of the formation of a Coulomb crystal.
Subsequently, the molecular beam valve was switched on to admit pulse trains of deflected \AP
molecules to collide and react with the spatially localized ions. Different parts of the deflected
molecular beam were directed at the stationary Coulomb crystal reaction target by tilting the
molecular beam setup. Ions that reacted with \AP formed product ions which remained trapped, but
were not laser cooled and, therefore, did not fluoresce~\cite{Willitsch:PRL100:043203}. These
product ions were sympathetically cooled by the remaining \Ca ions to form a dark shell around the
crystal. The progress of the reaction was monitored by observing the shrinking of the bright
fluorescing \Ca core of the Coulomb crystals as a function of time. Images of the crystals were
recorded every 30~s with a camera shutter time of 0.4~s over reaction times of typically 8 to
15~min. From the recorded images, the number of unreacted \Ca ions as a function of time was
determined from the crystal volumes~\cite{Willitsch:PRL100:043203}. Note that the \AP molecules in
the reaction volume were replenished with each gas pulse. Therefore, their number density was
essentially constant during the measurement time and the decrease of the number $N(t)$ of \Ca ions
in the crystal as a function of time $t$ followed pseudo-first-order kinetics. Pseudo-first-order
reaction rate constants $k_1$ were determined at specific deflection voltages and deflection
coordinates $y$ according to the rate law
\begin{equation}
   \ln\frac{N(y,t)}{N(y,t=0)}=-k_{1}(y)t.
\end{equation}
The deflection coordinate $y$ is defined as the offset of the deflected from the nominally
undeflected beam at the position of the Coulomb crystal. All measurements were performed with the
same power and detuning of the cooling laser from resonance to ensure a constant and well-defined
population of all three electronic levels of \Ca accessed during laser cooling (see
\autoref{fig:setup} (c)). The populations of the relevant \Ca states were determined from a
calibrated eight-level optical Bloch equation treatment including the effects of magnetic
fields~\cite{Hall:MP111:2020}.

Reactions with residual background H$_2$ gas in the ion trap chamber also contributed to a removal
of \Ca ions from the trap. The corresponding loss rates were measured for each set of experiments
following the same procedures described as above but without admitting the molecular beam. The
resulting values for the background loss rates were subtracted from the measured rates in the actual
reaction experiments. We note that collisions with the Ne carrier gas of the molecular beam did not
lead to any observable loss of \Ca ions from the trap, as confirmed by control experiments with pure
Ne beams.

\subsection{Mass spectrometry of trapped ions}
\label{sec:ms}
The ionic reaction products were analyzed using resonant-excitation (RE) mass spectrometry of the
Coulomb crystals~\cite{Willitsch:PCCP10:7200, Hall:MP111:2020}. Here, the motion of specific ion
species was resonantly excited by scanning the frequency of an additional RF drive voltage
($0.2$--$0.3$~V) applied to one of the trap electrodes. When the RF field was resonant with the
motional frequency of a trapped ion species, the Coulomb crystal heated up. This lead to a
dislocation of the \Ca ions from their equilibrium position. RE mass spectra were recorded by slowly
scanning the excitation frequency while monitoring the increase of the fluorescence yield in a
region close to but outside the normal extent of the Coulomb crystal. RE mass spectra of
multi-component crystals generally show broad peaks that are shifted with respect to single-species
crystals~\cite{Roth:PRA75:023402}. The exact intensity and position of the features depends on the
scan speed, the drive amplitude, the scan direction and the crystal composition. Therefore, RE mass
spectrometry only allows an approximate determination of the masses of the species present in a
multi-component Coulomb crystal~\cite{Hall:MP111:2020}.

\subsection{Molecular beam profile measurements}
Spatial deflection profiles of \AP were recorded in a time-of-flight (TOF) mass spectrometer that
replaced the ion-trap apparatus. \AP molecules were ionized via resonance-enhanced two-photon
ionization (R2PI) by a frequency-doubled pulsed dye laser pumped by a
Nd:YAG laser  with a repetition rate of 20~Hz. Pulses of 10~ns duration with
an energy of approximately $0.4~$mJ were focused to a spot size of 240~$\mu$m in the interaction
volume. The molecular ions were mass-selectively detected by their arrival time on a
multi-channel-plate (MCP) detector. \cis- and \trans-\AP were differentiated through their distinct
excitation wavenumbers of $34109~$cm$^{-1}$ and $34467~$cm$^{-1}$,
respectively~\cite{Reese:JACS126:11387}.

\section{Theoretical and computational methods}
\label{sec:theory:method}
\subsection{DFT calculations of reaction paths on the ground-state potential energy surface}
Short-range ion-molecule interactions were investigated computationally using density
functional theory (DFT) calculations. Stationary points along reaction paths to two possible
products were computed at the DFT~MPW1K/cc-pVTZ level of theory, using the Gaussian~09 software
suite~\cite{Gaussian:2009A02, Lynch:JPCA104:4811, Dunning:JCP90:1007}. Transition state structures
were calculated by a quadratic-singular-transit approach (QST-3)~\cite{Peng:IJC33:449,
   Peng:JCC17:49}, from energy-minimized \Ca-\AP and \Ca-product-radical complexes and an initial
transition state guess. To verify convergence to a saddle point, the resulting transition state
structure was distorted and resubmitted as a starting point in a new QST-3 calculation for the
transition state search. Basis set superposition errors were corrected using the counterpoise
routine provided in Gaussian~09.

\subsection{Adiabatic capture theory}
Long-range ion-molecule capture kinetics were modeled using the adiabatic capture theory
developed by Clary and co-workers~\cite{Clary:MP54:605, Stoecklin:JCSFT88:901}. The long-range
interaction potential $V$ between an ion and a polar molecule was approximated by the sum of the
dominant charge-permanent dipole\footnote{Here, ``permanent dipole'' refers to the dipole moment of
   \AP in its molecular frame~[\onlinecite{Klemperer:JPC97:2413}].} and charge-induced dipole
interactions
 \begin{equation}
    V(R,\beta)=-\frac{q\mu_D\cos\beta}{R^2}-\frac{q^2\alpha}{2R^4},
   \label{eqn:v}
\end{equation}
where $R$ is the distance between the ion and the center of mass of the \AP molecule, $\mu_D$ is its
permanent electric dipole moment, $\beta$ the orientation angle of the molecular
dipole moment with the ion-molecule axis, $q$ the charge of the ion and $\alpha$ the scalar
polarizability of \AP. Using the methods described in ref.~\onlinecite{Stoecklin:JCSFT88:901},
centrifugally corrected and rotationally adiabatic potential energy curves for the system
$\Ca+\cis$\/-/\trans-\AP were calculated for \AP rotational states with quantum numbers ranging from
$j=0$ to $j=100$ for $R$ between 2 and 48~\bohr. The dipole moments of \AP were taken from
ref.~\onlinecite{Filsinger:PCCP10:666}, the isotropic polarizabilities were calculated at
the DFT B3LYP/aug-cc-pVTZ level of theory. Rotational-state-specific reaction cross sections for
$j=0$ up to $j=15$ were calculated from a summation over all partial waves for which the maximum of
the centrifugally corrected potential energy curve did not exceed the experimental collision energy.
Effective capture rate constants were calculated by multiplying the state-specific cross sections
with the velocity  and the relevant state populations  at the rotational
temperature of \AP in the molecular beam.

\subsection{Molecular-dynamics simulations of Coulomb crystals}
\label{sec:md}
Fluorescence images and RE mass spectra of the multi-component Coulomb crystals were simulated using
molecular dynamics (MD) methods. MD simulations were performed using a modified version of the
Protomol program package~\cite{Matthey:ATMS30:237}. Fluorescence images were simulated from ion
trajectories calculated by solving the classical three-dimensional equations of motion of the ions
in the trap under the influence of laser cooling~\cite{Willitsch:PCCP10:7200,Bell:FD142:73}. To
minimize computer time, an isotropic friction force to emulate laser cooling and the pseudopotential
approximation for the ion trap was used~\cite{Willitsch:IRPC31:175}. RE mass spectra were simulated
following the methods described in Ref.~\onlinecite{Hall:MP111:2020, Roth:PRA75:023402}. Briefly,
the Coulomb crystals were offset from the central trap axis by 20~$\mu$m at the beginning of the
simulation and allowed to relax. The Fourier transform of the time-dependent total kinetic energy of
the ions yielded the frequency spectrum of the Coulomb crystal which is also the experimental
observable. The frequencies obtained by this method were calibrated using a comparison of a measured
and calculated RE mass spectrum of a pure \Ca crystal. A correction factor of 0.92 was applied to
the calculated frequencies of each spectrum to achieve optimal agreement with the experiment.

\subsection{Monte-Carlo simulations of molecular beam profiles}

\begin{figure}
   \centering
   \includegraphics[width=\linewidth]{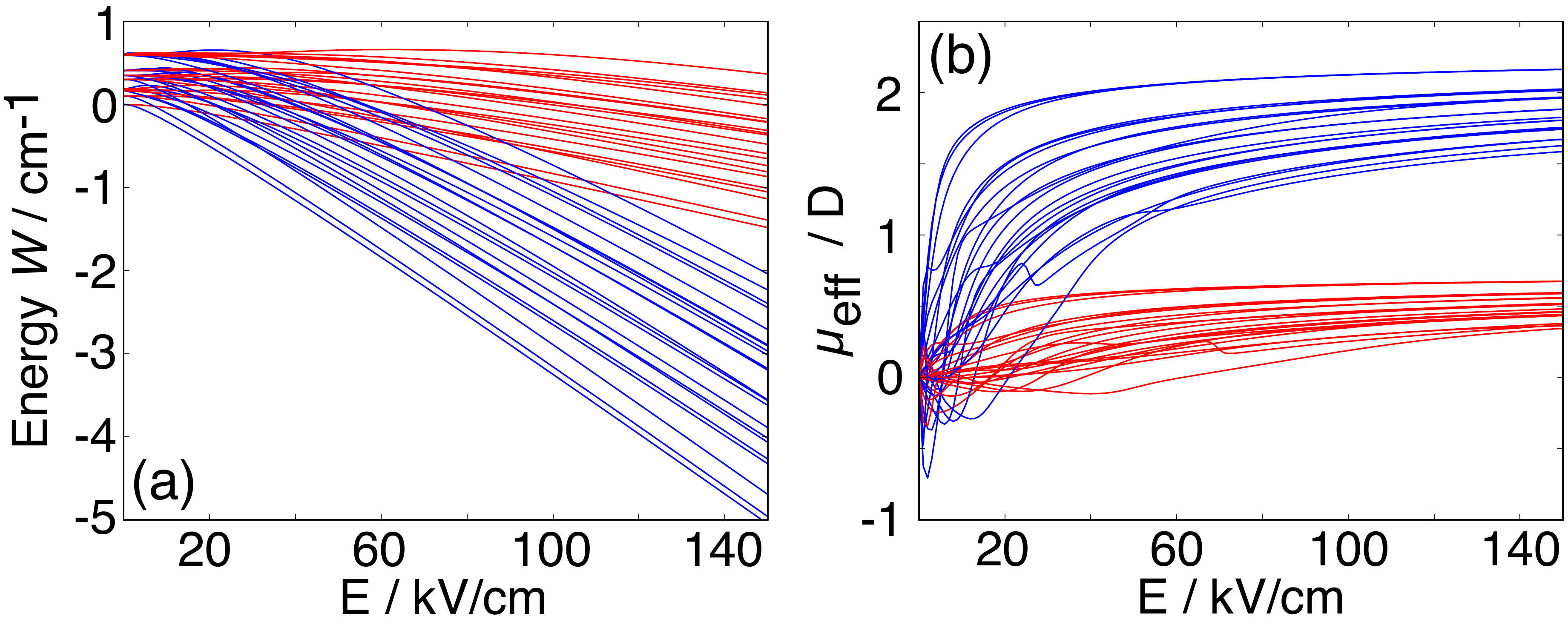}
   \caption{(a) Stark energies $W$ for the lowest rotational quantum states $j=0$--$2$ of (blue)
      \cis- and (red) \trans-\AP as a function of the electric field strength $E$. (b) Effective
      dipole moments $\mu_{\text{eff}}$ for $j=0-2$ of \cis- (blue) and \trans- (red) \AP.}
   \label{fig:stark_curves}
\end{figure}

The simulation of spatial deflection profiles has been described in detail
previously~\cite{Kuepper:FD142:155, Filsinger:JCP131:064309}. Briefly, the electric field $E$ and
its gradient ($\vec{\nabla}E$) were calculated using finite element methods implemented in the
COMSOL Multiphysics program. Stark energy curves $W(E)$ of the \AP quantum states and their
effective dipole moments $\mu_{\text{eff}}$ were calculated using CMIstark~\cite{Chang:CPC185:339}
(see \autoref{fig:stark_curves}). From the electric fields and Stark energy curves, the molecular
beam deflection profiles were calculated with libcoldmol~\cite{Filsinger:JCP131:064309}.
Trajectories for molecules in individual rotational quantum states were obtained by numerical
integration of the 3D equations of motion using a Runge-Kutta algorithm. The initial conditions
according to the parameters of the molecular beam were sampled by a Monte-Carlo approach, and every
individual molecule was propagated through a simulated beamline that includes all mechanical
apertures of the experimental setup.

The spatial deflection profile $I(y,\Trot)$ for an ensemble of molecules at a rotational temperature
$\Trot$ was calculated from the single-quantum-state deflection profiles $I_s(y)$ using
\begin{equation}
   I(y,\Trot)=\frac{1}{w}\sum^{N}_{s=1}w_s(\Trot)I_s(y).
\end{equation}
Here, $N$ is the number of quantum states included in the simulation and
\mbox{$w_s(\Trot)=g_Mg_{\text{ns}}e^{(W_0-W_s)/(k_\text{B}\Trot)}$} is the population weight for a
given quantum state. $W_0$ is the field-free energy of the ground state and $W_s$ the field-free
energy of state $s$. $g_M=1$ for $M=0$ and $g_M=2$ otherwise. $g_{\text{ns}}$ accounts for nuclear
spin statistical weight of the current state with $g_{\text{ns}}=1$ for all rotational states of
\AP. The normalization constant is given by $w=\sum\limits^N_{s=1}w_s$.

\section{Experimental  Results}
\label{sec:exp_result}
\subsection{Deflection curves of \AP}
\label{sec:result:deflection}
The density of each conformer in the deflected and dispersed molecular beam was measured by
recording the number of R2PI-ionized \cis and \trans conformers of \AP as a function of the
deflection coordinate $y$.
\begin{figure}[t]
   \centering
   \includegraphics[width=\linewidth]{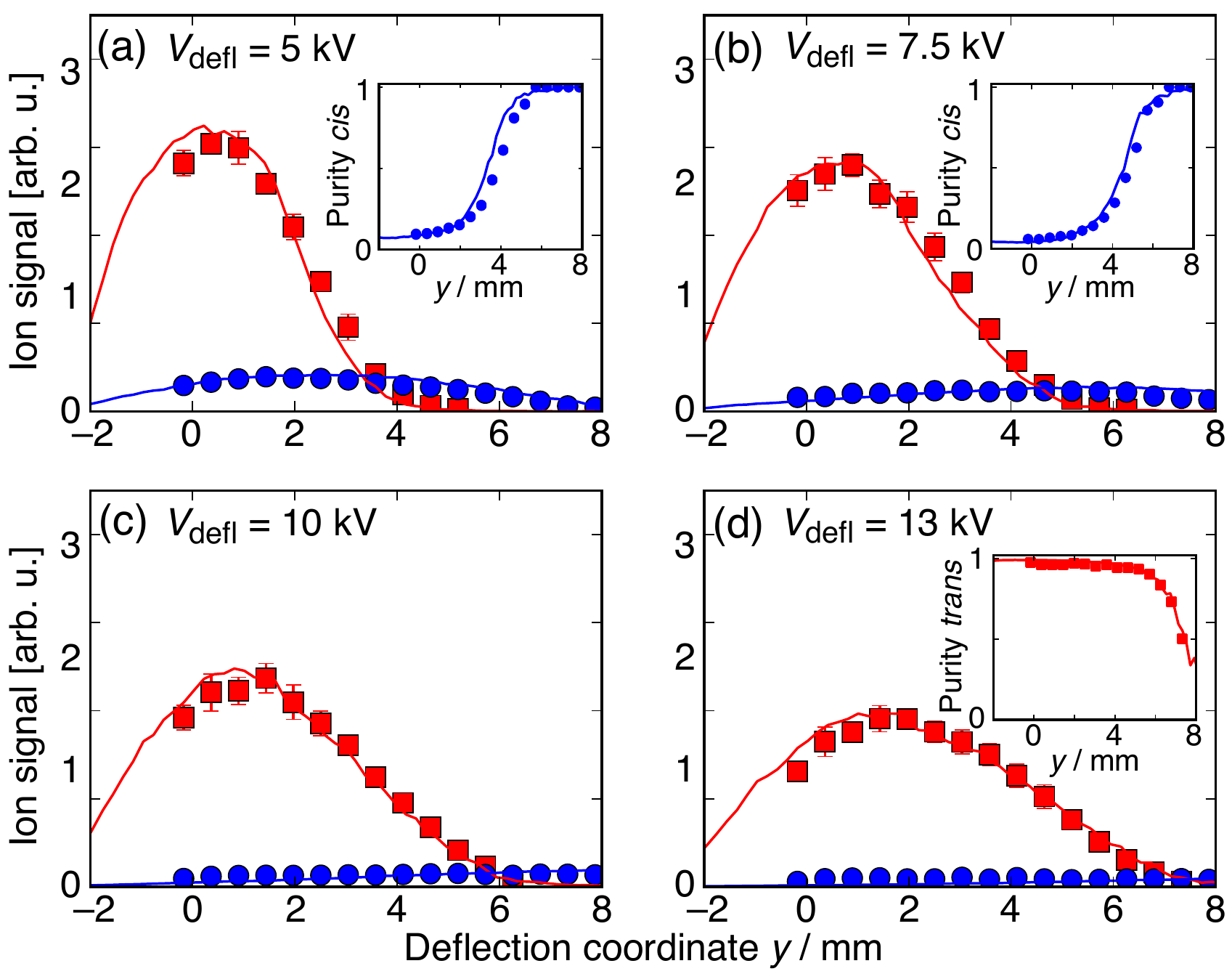}
   \caption{Density profiles of the deflected beam of \cis (blue) and \trans (red) \AP at deflector
      voltages $V_\text{defl}=$ (a) $5$~kV, (b) $7.5$~kV, (c) $10$~kV and (d) $13$~kV. These data
      were measured by conformer-specific multi-photon ionization as a function of the
      molecular-beam deflection coordinate $y$. Solid lines: corresponding Monte-Carlo trajectory
      simulations. In the insets, purities of (b) \cis and (d) \trans conformers are given, as
      obtained by dividing the relevant conformer density profile by the sum of the \cis and \trans
      profiles. Error bars indicate the statistical $95\,\%$ confidence interval of the data
      points.}
   \label{fig:defl:exp-results}
\end{figure}
The measured conformer-selective deflection profiles are shown in \autoref{fig:defl:exp-results}, in
which each data point represents the signal averaged over 1000 laser shots. When high voltages were
applied to the deflector, both conformers were deflected upwards. The deflection was considerably
larger for the more polar \cis-\AP. For instance, for a deflector voltage of 7.5~kV above $y=6$~mm a
pure sample of \cis conformers was obtained (see \autoref{fig:defl:exp-results}\,(a)). The insets in
\autoref{fig:defl:exp-results} show that the fraction of \cis-\AP in the probed sample can be
continuously tuned as a function of $y$. At heights above the cut-off of the \trans-\AP beam
profile, the density of the \cis conformers is still comparable to its density in the free jet, \ie,
it is only decreased to one fourth. When increasing the voltages to 13~kV
(\autoref{fig:defl:exp-results}\,(d)), \cis-\AP was deflected so strongly that it was essentially
depleted from the detection region. As a consequence, an almost clean sample of \trans-\AP was
obtained.

Monte Carlo simulations of the deflection curves are shown as solid lines in
\autoref{fig:defl:exp-results}. The simulations at an initial rotational temperature of $1.1$~K
agree well with the experimental profiles. In particular, the fractional intensities plotted in the
insets were reproduced by the simulations (solid lines). The simulated profiles and the measured
population ratios of the two conformers were used for fitting conformer-specific rate constants from
measured reaction rate profiles as described in the following sections.

\begin{figure}[t]
   \centering
   \includegraphics[width=\linewidth]{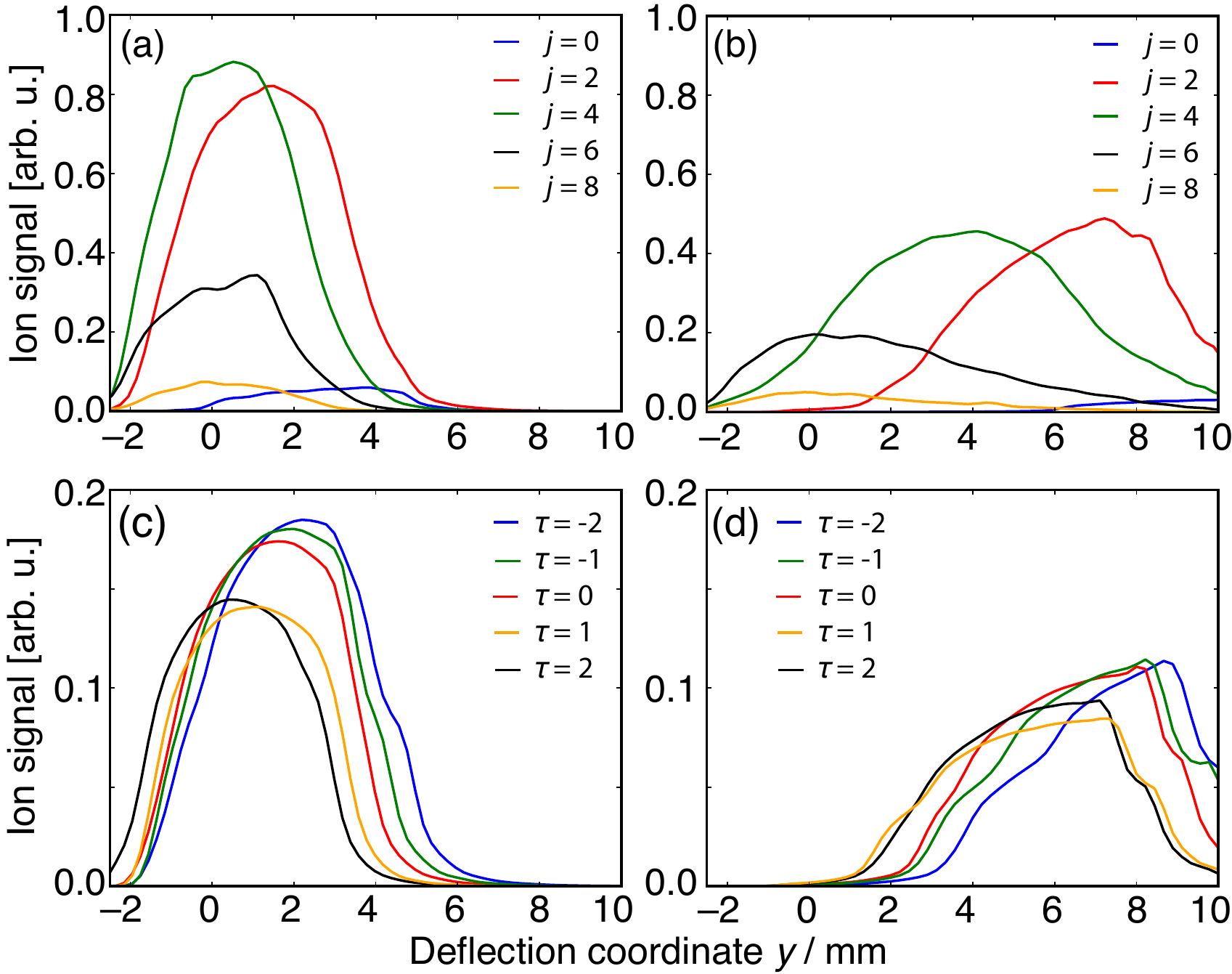}
   \caption{Deflection profile simulations at $V_\text{defl}=7.5$~kV for even $j=0$ to $j=8$ for
      specific rotational states $j$ of (a) \trans- and (b) \cis-\AP, and for specific
      asymmetric-top quantum numbers $\tau$ at $j=2$ of (c) \trans- and (d) \cis-\AP. See text for
      details.}
   \label{fig:defl:simulation}
\end{figure}
For the simulated deflection profiles shown in \autoref{fig:defl:exp-results}, different rotational
states $j$ have different spatial distributions according to their different effective dipole
moments \mueff. \autoref{fig:defl:simulation} shows the simulated deflection profiles of individual
states ranging from $j=0$ to $j=8$. For each profile, the contribution of all $j_\tau$ states
weighed with their relative thermal populations at the rotational temperature of 1.1~K and their
statistical weights were included. Thus, the area underneath each $j$ profile in
\autoref{fig:defl:simulation} represents the relative thermal population of each $j$ manifold at
1.1~K as well as their relative contribution to the reactions. Comparing the deflection profiles of
individual $j$ states for the two conformers, profiles of low $j$ of \cis-\AP exhibit significantly
stronger spatial deflection than those of \trans-\AP. However, for high $j$ states both conformers
show similar deflection patterns demonstrating the quickly vanishing dipole moment for rotationally
excited species and the need for very cold molecular beams~\cite{Kierspel:CPL591:130}.

\subsection{Number density of \AP}
\label{sec:nd}
The absolute number density of \AP in the molecular beam was derived by calibration against the
reaction $\NNO+\Ca\rightarrow\text{CaO}^++\text{N}_2$. From measurements of the pseudo-first-order
rate constants for this reaction and the reported value for the second-order-rate
constant~\cite{Plane:JPCA110:7874}, the density of \NNO molecules in the beam was determined. The
ratio of the experimentally determined first order rate constant to the known second order rate
constant was equivalent to the time-averaged number density $n_\text{avg}$. For
\mbox{$k_1=4.30(4)\times10^{-3}$~s$^{-1}$}, measured with a beam of 50~mbar of \NNO seeded in 30~bar
of Ne we obtained $n_\text{avg}(\NNO)=3.84(62)\times10^{7}~\text{cm}^{-3}$. Assuming that the number
densities in the beam were proportional to the partial pressures before expansion (for \AP at
$145~\degree$C approximately 10~mbar~[\onlinecite{Chickos:JPCRD31:537}]), the \AP density was
estimated to be \mbox{$n_\text{avg}=7.7(12)\times10^{6}$~cm$^{-3}$}.\footnote{For \AP, the number
   density in each gas pulse $n_\text{pulse}$ is equal to $2.56(41)\times10^{8}$~cm$^{-3}$.
   $n_\text{avg}$ equals to a product of the gas pulse duration (50~$\mu$s), the repetition rate
   (600~Hz), and $n_\text{pulse}$.}

\subsection{Reaction profiles and conformer-specific rate constants of \Ca+\,\AP}
\label{sec:result:reaction}
\begin{figure}
   \centering
   \includegraphics[width=\linewidth]{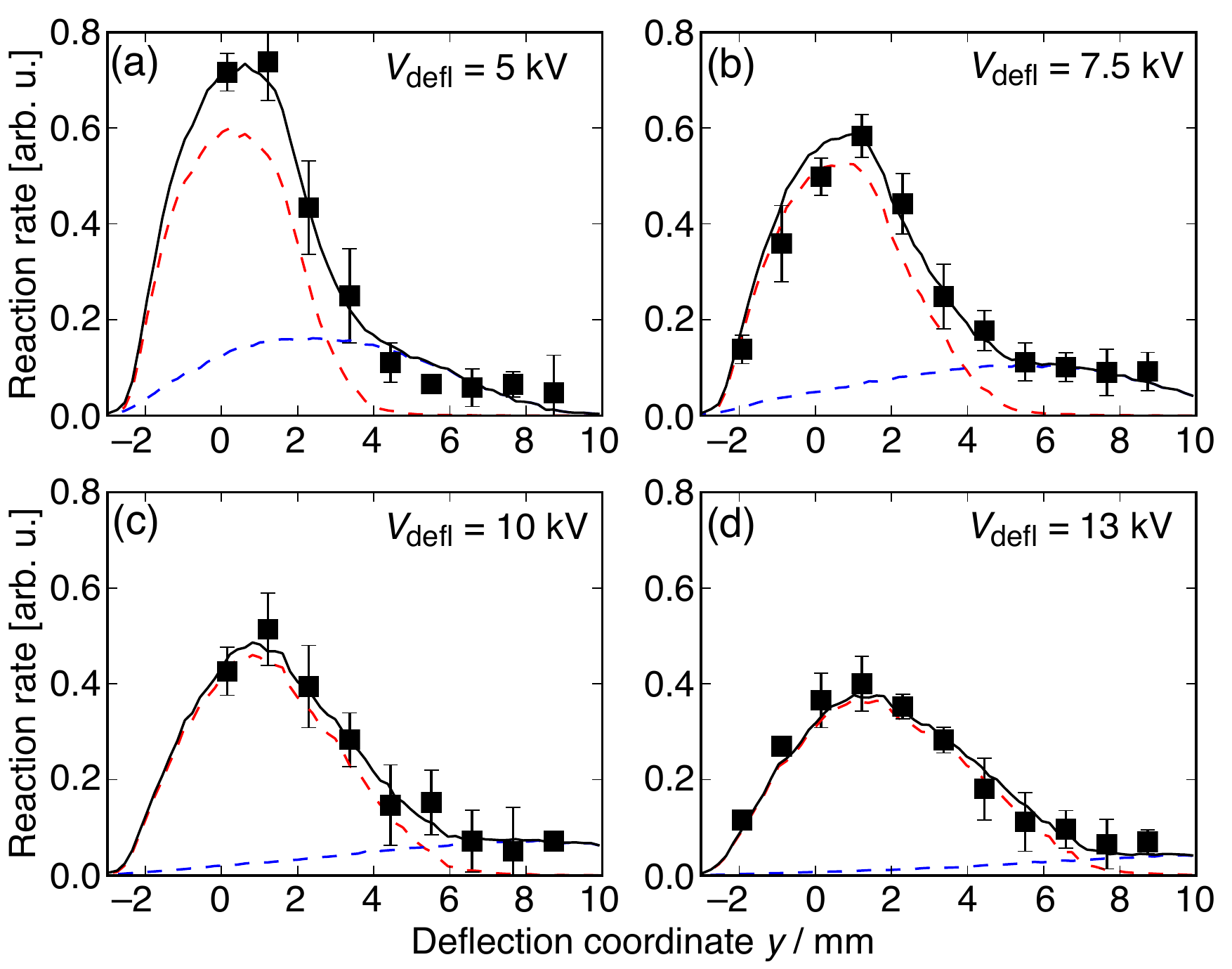}
   \caption{Reaction profiles (symbols) and their fits (lines) at (a) 5~kV, (b) 7.5~kV, (c) 10~kV
      and (d) 13~kV. The solid black lines represent the calculated total contributions of both
      conformers. Dashed lines represent individual contributions of the \trans (red) and \cis
      (blue) conformers. Error bars indicate the statistical $95\,\%$ confidence interval of the
      data points.}
   \label{fig:react_results}
\end{figure}
In \autoref{fig:react_results}, the experimentally determined pseudo-first-order rate constants
$k_{1,\text{total}}$ are shown as a function of deflection coordinate $y$ for four deflector
voltages $V_\text{defl}=5$, $7.5$, $10$ and $13$~kV. Each data point in \autoref{fig:react_results}
represents the mean of at least four individual reaction measurements. The measured rate constants
$k_{1,\text{total}}(y)$ reflect both, the density distributions of conformers in the deflected
molecular beam $n_{\trans/\cis}$ and the conformer-specific second-order rate constants
$k_{2,\trans/\cis}$ of the reaction:
\begin{equation}
   k_{1,\text{total}}(y)=k_{2,{\cis}}~n_{\cis}(y)+k_{2,{\trans}}~n_{\trans}(y)
   \label{eqn:rates}
\end{equation}
$n_{\trans}(y)$ and $n_{\cis}(y)$ were determined as described in
sections~\ref{sec:result:deflection} and \ref{sec:nd}. $k_{2,{\cis}}$ and $k_{2,{trans}}$ were
determined from a global fit of \eqref{eqn:rates} to the reaction-rate profiles in
\autoref{fig:react_results}. The fit yielded the conformer-specific rate constants
$k_{2,{\cis}}=2.3(9)\times10^{-10}~\qcms$, $k_{2,{\trans}}=1.1(4)\times10^{-10}~\qcms$ and the ratio
$k_{2,{\cis}}/k_{2,{\trans}}=2.1(5)$ within a 95\% confidence interval. This fit also yielded the
rotational temperature of the molecules in the beam to be 1.1~K.

\subsection{Variation of \Ca electronic state populations}
\label{sec:exp_result:detuning}
\begin{figure}[t]
   \centering%
   \includegraphics[width=\figwidth]{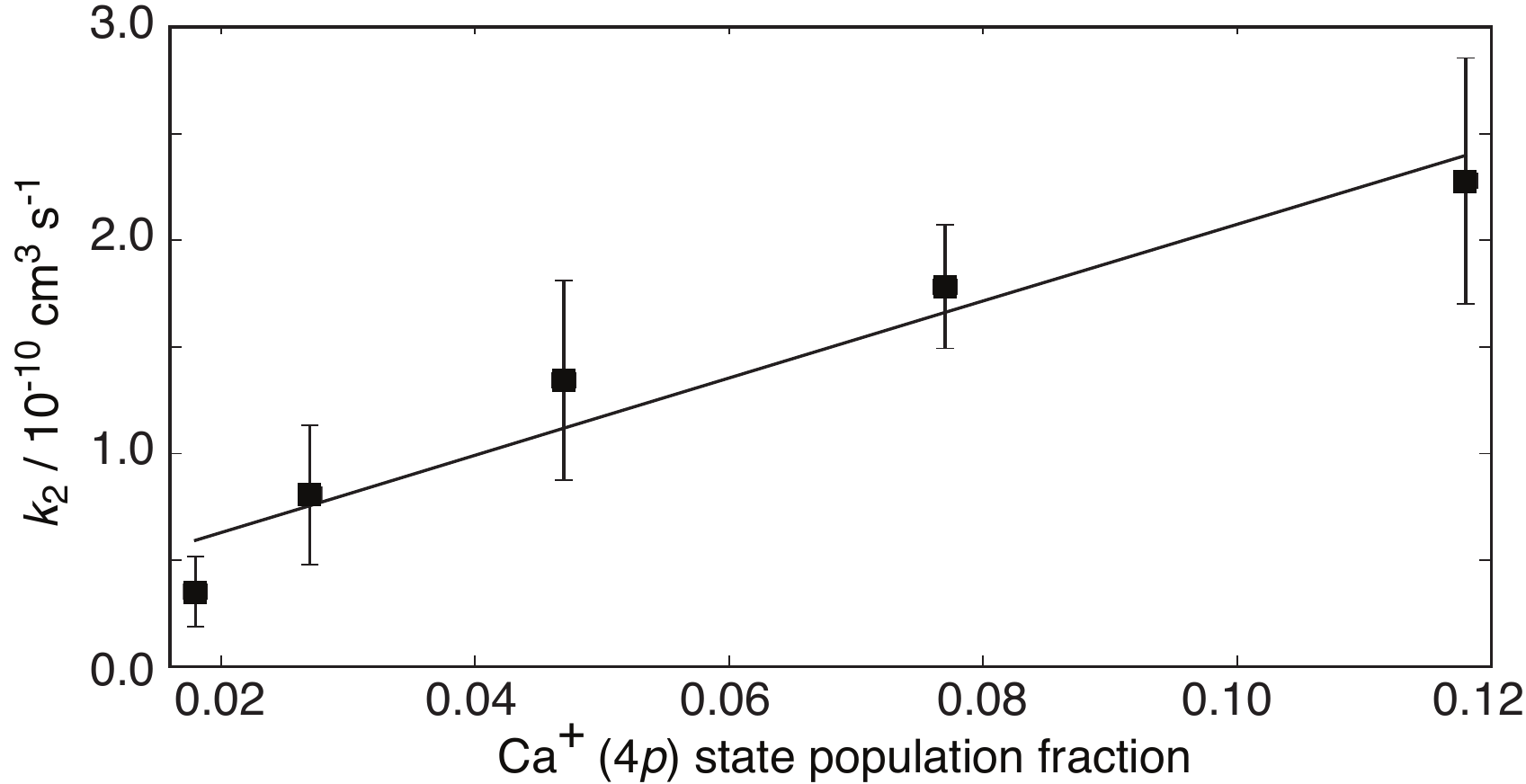}
   \caption{Conformer-averaged bimolecular rate constant $k_2$ as a function of the population in
      the \Ca $(4p)$ state. Error bars represent the statistical 95\% confidence interval. The line
      represents a linear regression to the data.}
   \label{fig:detuning:results}
\end{figure}
As described in \autoref{sec:setup:iontrap}, \Ca ions were constantly excited during laser cooling
so that collisions occurred between \AP and \Ca in the $(4s)~^2S_{1/2}$, $(3d)~^2D_{3/2}$ and
$(4p)~^2P_{1/2}$ states. To study the effect of the electronic excitation of \Ca on the reaction
rates, we varied the \Ca state populations by changing the detuning of the cooling laser beam from
the $(4s)\rightarrow (4p)$ resonance while optimizing the $(3d)\rightarrow (4p)$ repumping laser
detuning to achieve the best cooling conditions. \autoref{fig:detuning:results} shows the measured
rate constants as a function of the $(4p)$ state population using an undeflected molecular beam of
\AP molecules.
\begin{table}[b]
   \centering
   \caption{Bimolecular rate constants $k_2$ for reactions of \Ca in its relevant electronic states
      with \AP molecules in an undeflected beam.}
   \bigskip
   \begin{tabular}{c|c}\hline
      Reaction channel & $k_2~/~\text{cm}^3~\text{s}^{-1}$ \\ \hline
      Ca$^+~(4s)~^2S_{1/2}$~+~\AP & $2.66(44)\times10^{-11}$ \\
      Ca$^+~(3d)~^2D_{3/2}$~+~\AP & $2.69(45)\times10^{-12}$ \\
      Ca$^+~(4p)~^2P_{1/2}$~+~\AP & $1.91(32)\times10^{-9}$ \\ \hline
   \end{tabular}
   \label{tab:rate-constants}
\end{table}
From this set of measurements, the state-specific rate constants $k_2$ listed in
\autoref{tab:rate-constants} were derived following the procedures outlined in
Refs.~\onlinecite{Hall:PRL107:2432002, Hall:MP111:2020}. The rate constant for reactions out of the
excited $(4p)$ state was found to be two to three orders of magnitude larger than the rate
coefficients for reactions out of the $(4s)$ and $(3d)$ states. In the experiments reported in this
paper, a detuning of 47-56~MHz was used, yielding a population of the $(4p)$ level of $5-10\%$.

Because of its large rate constant, this channel dominates the reaction rates observed in the
experiment and the contribution of the other channels can be neglected in good approximation. Thus,
one can assume that the conformer-specific rate constants determined in
\autoref{sec:result:reaction} only reflect reactions with \Ca $(4p)$. Scaled to a state population
of 100\%, the conformer-specific second-order rate-constants
$k_{2,{\cis}}=3.2(13)\times10^{-9}~\qcms$ and $k_{2,{\trans}}=1.5(6)\times10^{-9}~\qcms$ for the
reaction of \cis-\AP and \trans-\AP, respectively, with Ca$^+$ $(4p)$ were obtained.

\subsection{Mass spectra of reaction products}
\begin{figure}[t]
   \centering%
   \includegraphics[width=\figwidth]{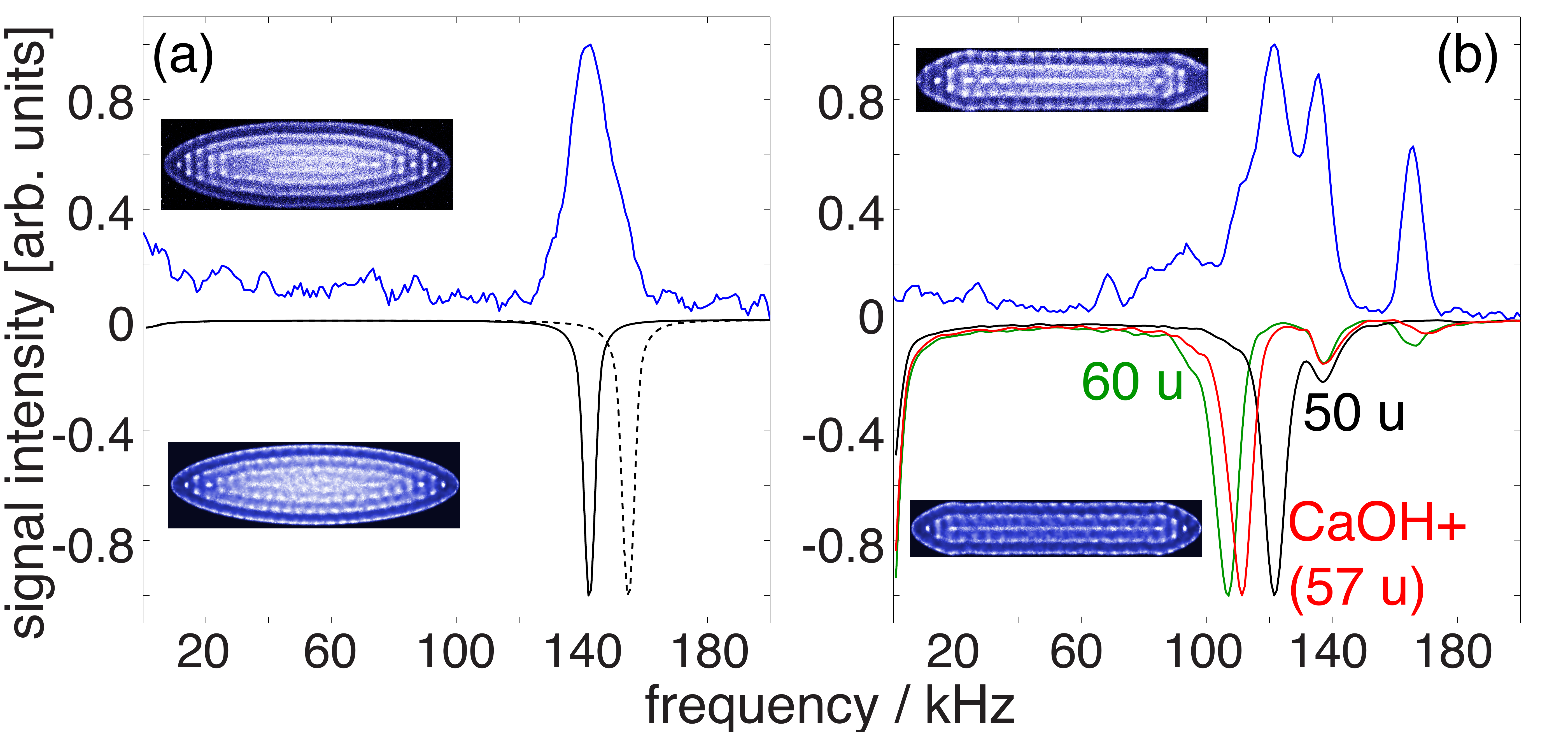}
   \caption{Resonance-excitation mass spectra (upper traces) and their molecular dynamics
      simulations (lower inverted traces). (a) Top: experimental spectrum of a pure \Ca crystal.
      Bottom, dashed line: corresponding  simulation using a crystal with 675 ons. Solid line:
      simulated spectrum scaled by 0.92 along the frequency axis to match the experiment. (b) Top:
      experimental spectra after a reaction time of 8~min with \AP. Bottom: scaled simulated spectra
      of crystals composed of 350 \Ca and 325 heavy ions with mass 50~u (black), 57~u (red), and
      60~u (green).}
   \label{fig:REMSresults}
\end{figure}
\autoref{fig:REMSresults}\,(a) and (b) show RE mass spectra before and after a typical reaction,
respectively. In the spectrum of the pure \Ca Coulomb crystal, \autoref{fig:REMSresults}\,(a), a
single peak at an excitation frequency of $140$~kHz was observed. This feature was also present in
the RE mass spectrum of the multi component crystal after the reaction and could unambiguously be
assigned to the excitation of Ca$^+$ ions with a mass of 40~u. The spectrum of the Coulomb crystal
after reaction showed two additional strong peaks at $120$ and $165$~kHz. The feature at lower
frequency was assigned to product ions. MD simulations for crystals composed of 350 \Ca and 325
heavier ions indicate that the product ion mass is in the range of 50 to 60~u, suggesting that the
reaction products are CaOH$^+$ (57~u) or CaNH$_2^+$ (56~u). As discussed in detail in Refs.
\cite{Roth:PRA75:023402, Hall:MP111:2020}, our approximate simulation approach cannot be expected to
perfectly reproduce the observed peak positions and intensities in the spectra as it is not a
faithful representation of the complex processes leading to the signal measured in the experiments
(see \autoref{sec:md}). Nonetheless, the MD simulations serve as a useful guide for the
interpretation of the mass spectra.

Based on the analysis of the ion trajectories obtained in the MD simulations, the peak at
$\approx165$~kHz was assigned to a high-frequency excitation of \Ca ions in the combined potential
of the trapping fields and the product ions. The weak broad signal in the range from $60$--$100$~kHz
is indicative of the presence of even higher masses, possibly arising from consecutive reactions of
the primary product ions with \AP from the molecular beam.

\section{Reaction mechanisms and kinetics}
\label{sec:theory:results}
\subsection{Reaction pathways on the ground-state potential energy surface}
According to \autoref{tab:rate-constants}, the \Ca $(4p)$ state rate constant is at least two orders
of magnitude larger than those in the $(4s)$ and $(3d)$ states, suggesting different reaction
dynamics for these channels. The large values for the rate constants obtained for reactions in the
\Ca $(4p)$ state are indicative of a capture process~\cite{Rowe:IJMSIP80:239, Clary:ARPC41:61}. In
this case, the reaction rate is limited by the rate of formation of the reaction complex. Afterward,
the reaction proceeds with near unit efficiency. The kinetics of the reaction are then solely
controlled by long-range intermolecular interactions. For reactions with \Ca~$(4s)$ and $(3d)$,
however, the significantly smaller rate constants compared to the capture limit are indicative of
the existence of barriers on the reaction path which limit the reaction
rates~\cite{Sabbah:Science317:102, Gingell:JCP133:194302}. In the next paragraphs, the possible
roles of the different \AP conformations in these two types of situations are discussed.

While a high-level \emph{ab initio} calculation for the potential energy surfaces of the excited
channels is beyond the scope of the present work, the results of the DFT calculations for the
ground-state surface can nonetheless give valuable insights into possible transition-state (TS)
structures and reaction pathways.
\begin{figure*}[t]
   \includegraphics[width=\textwidth]{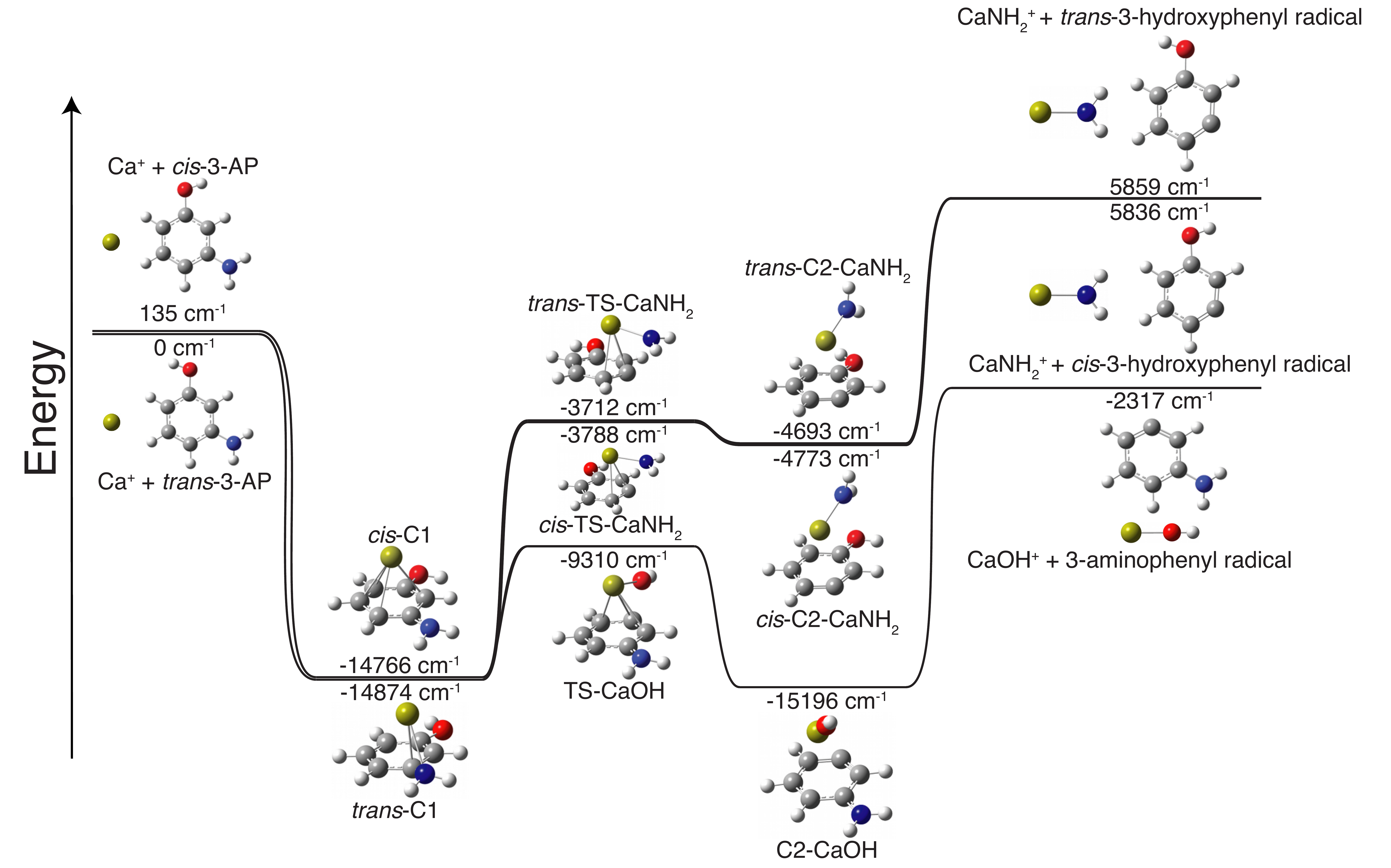}
   \caption{\sloppy Schematic energy diagram of stationary points and transition states on the
      \mbox{$\Ca(4s)+\cis/\trans$-3AP} potential energy surface. See text for details.}
   \label{fig:ped}
\end{figure*}
\autoref{fig:ped} shows a schematic potential energy diagram of stationary points and TS of the
\mbox{$\Ca(4s)+\cis/\trans$-\AP} reaction. The reactants form the ion-molecule complex C1 in the
entrance channel in which \Ca is bound above the aromatic ring. From C1, the reaction proceeds
either by abstraction of OH or NH$_2$, yielding the ionic products CaOH$^+$ and CaNH$_2^+$,
respectively. For the pathway leading to CaOH$^++~$3-aminophenyl radical, the reaction proceeds
through TS-CaOH. This TS was found to be identical for both conformers of \AP as the OH group is
displaced from the aromatic ring towards the \Ca ion. The resulting products are identical for both
conformers. This pathway is calculated to be exothermic by $\approx0.3$~eV (2320~cm$^{-1}$).

For the second pathway, leading to CaNH$_2^+$ and \cis/\trans-3-hydroxyphenyl radical, a TS
structure TS-CaNH$_2$, analogous to TS-CaOH, was found. The \Ca ion is coordinated above the
aromatic ring and the amino group is displaced out of the plane towards the ion. Two different
TS-structures for the two conformers of \AP exist. They differ by the orientation of the OH-group
with respect to the amino group. Their energy difference was calculated to be 76~cm$^{-1}$. The
conformational dependence is preserved throughout the product channel, but the energy difference
between the two conformeric pathways is very small. The pathway leading to CaNH$_2^+$ is calculated
to be endothermic by $\approx0.7$~eV ($5646$~cm$^{-1}$). Under the present conditions, this second
pathway is expected to be thermodynamically accessible only for reactions with \Ca in the excited
$(4p)$ and $(3d)$ states.

The present DFT calculations predict that \Ca in its ground state reacts with \AP to CaOH$^{+}$ via
a submerged transition state, TS-CaOH in \autoref{fig:ped}, that is lower in energy than the
reagents. The energy profile along the reaction coordinates is reminiscent of the situation in
related abstraction reactions, \eg, \Ca + CH$_3$F~\cite{Harvey:CPL273:164, Willitsch:PRL100:043203}
or the reactions of O atoms with alkenes~\cite{Smith:FD133:137, Sabbah:Science317:102}. The present
calculations predict the TS structure to be 1.2~eV lower in energy than the reactants. Comparing
with the kinetics in similar type of reactions as observed in Refs.~\onlinecite{Harvey:CPL273:164,
   Smith:FD133:137, Sabbah:Science317:102}, it is difficult to see how the low-lying barrier in the
present case can lead to a rate constant for the \Ca~$(4s)$ + \AP channel about two orders of
magnitude lower than the capture limit observed (see \autoref{sec:exp_result:detuning}). Possible
reasons for this discrepancy could be an underestimation of the barrier height by the current DFT
approach or the existence of additional barriers or dynamic constraints that have not been accounted
for. Overall, the present calculations give no indication of a short-range reaction mechanism that
could explain the factor of 2 difference in the observed rate constants for the reaction of
\cis/\trans-\AP with \Ca in its ground state.

\subsection{Capture dynamics in the \Ca$(4p)$+\AP excited channel}

\begin{figure}[t]
   \centering
   \includegraphics[width=0.5\textwidth]{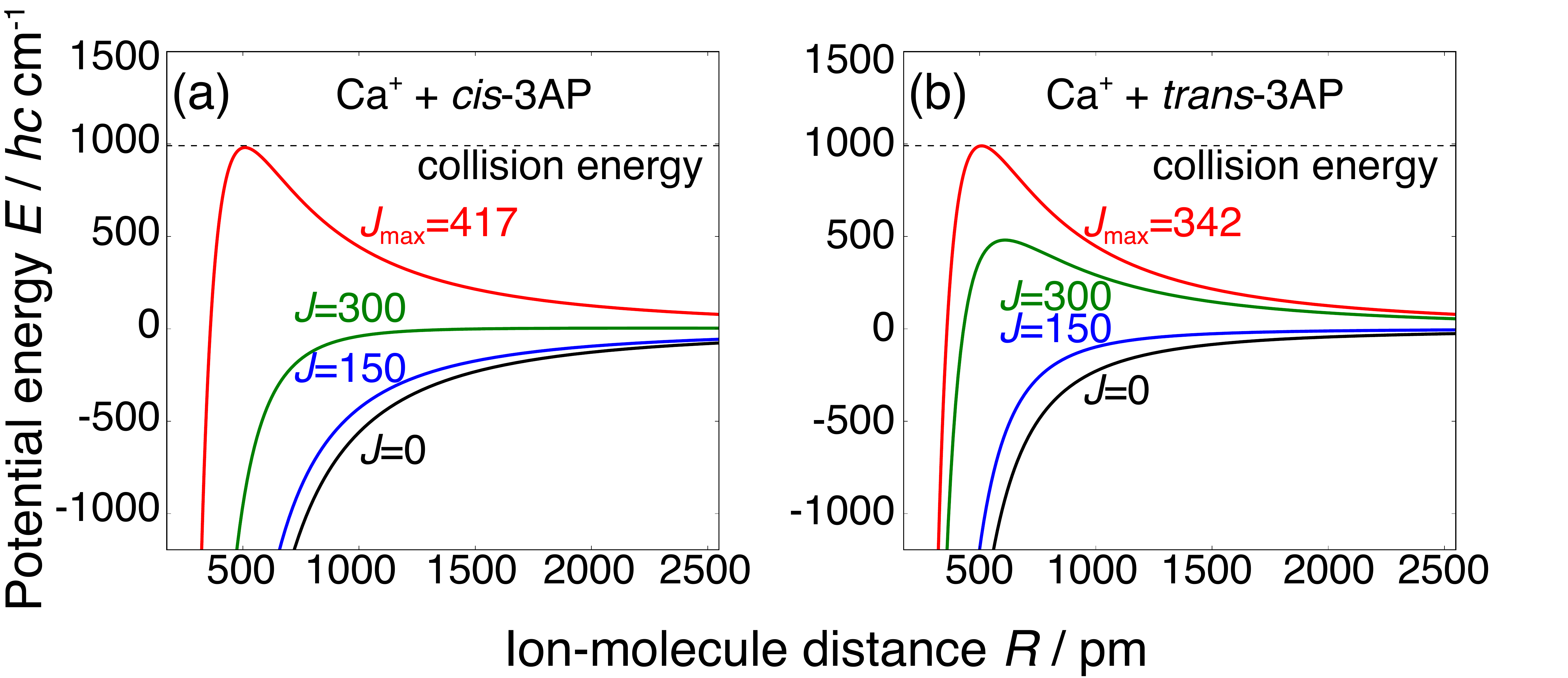}
   \caption{Centrifugally corrected long-range interaction potentials for (a) \cis- and (b)
      \trans-\AP in the rotational ground state $j=0$. From Ref. [\onlinecite{Chang:Science342:98}]. Reprinted with permission from AAAS.}
   \label{fig:capturecurves}
\end{figure}
\autoref{fig:capturecurves} shows centrifugally corrected adiabatic potential energy curves for the
reaction of \Ca$(4p)$ with \cis and \trans-\AP in their $j=0$ rotational states. In the case of
\cis-\AP, the centrifugal barrier is more strongly suppressed and reactive collisions proceed up to
larger maximum values $J_{max}$ of the total angular momentum. For the collision energy of the
present study (0.123~eV), we find $J_{max}$ = 417 and 342 for the \cis- and \trans-conformers,
respectively. Thus, in a classical picture, a larger impact parameter $b_{max}=J_{max}/\mu v$
results for the \cis-conformer, with $\mu$ being the reduced mass and $v$ the collision velocity so
that a larger reaction cross section $\sigma=\pi{}b^2_{max}$ is obtained for the \cis compared to
the \trans species.

\begin{figure}
   \centering
   \includegraphics[width=0.5\figwidth]{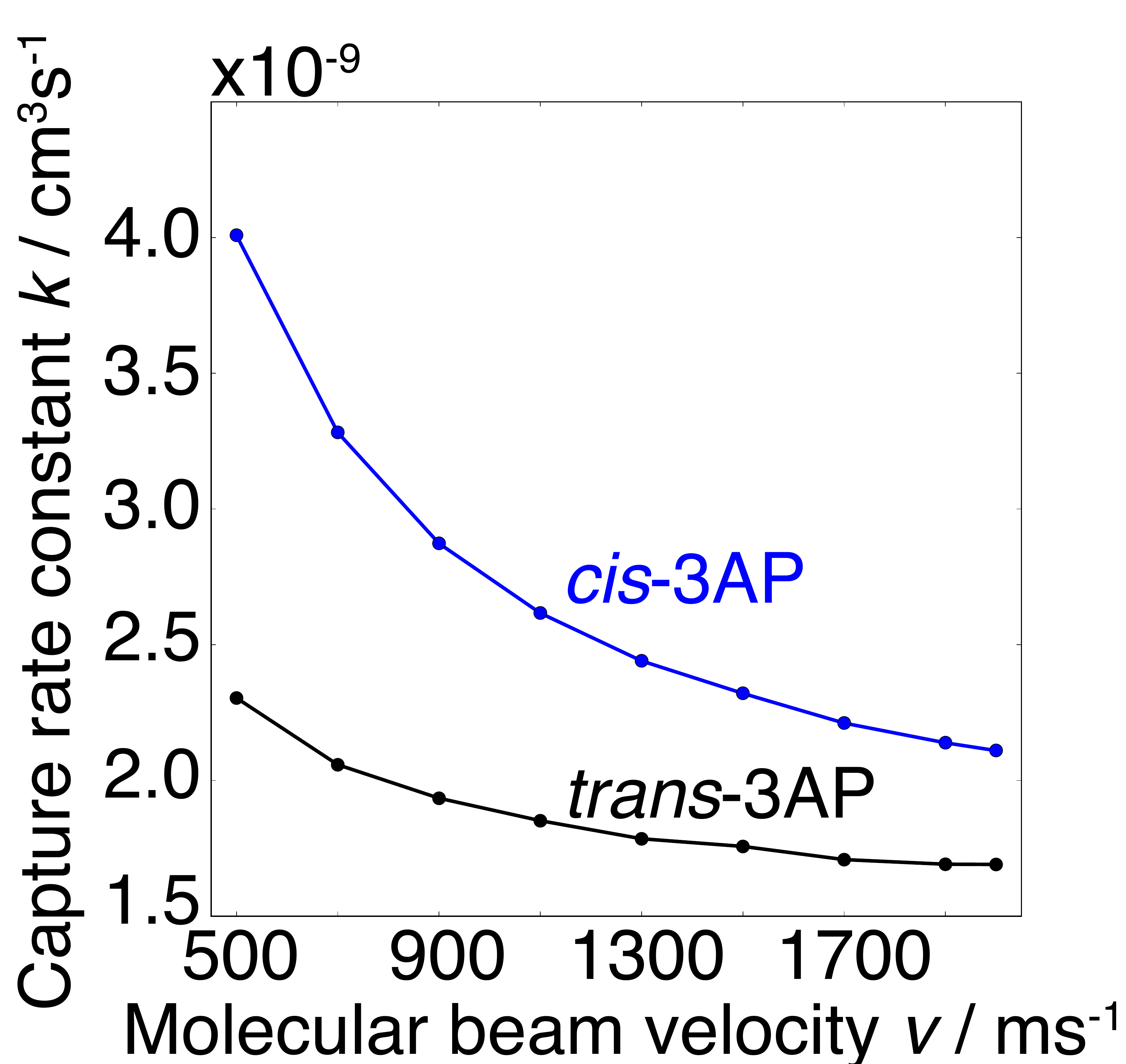}
   \caption{Capture rate constants of \cis- and \trans-\AP in the rotational quantum state $j=0$ as
      a function of the collision velocity. The experimental collision velocity amounted to
      $v=900$~ms$^{-1}$.}
   \label{fig:captureecol}
\end{figure}
As shown in \autoref{fig:captureecol}, the calculated capture rate constants depend on the collision
energy. Moreover, the monotonic increase of rate constants with decreasing collision energy is more
prominent for the \cis species with the larger dipole moment. Therefore, the ratio of the reactivity
between the two conformers becomes larger for smaller collisional energies.
\begin{figure}
   \centering
   \includegraphics[width=\figwidth]{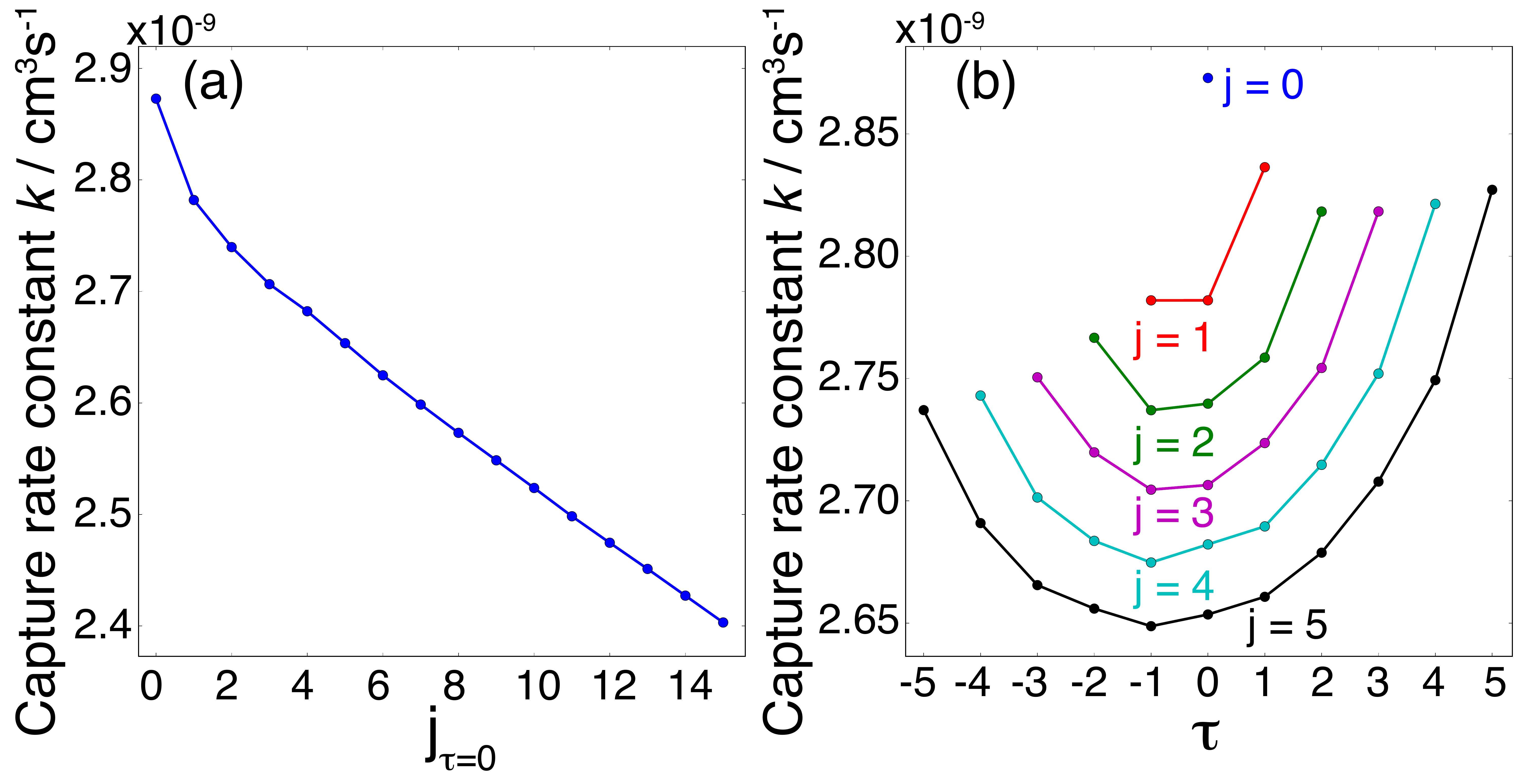}
   \caption{State-dependent capture rate constants $k$ for \cis-3-aminophenol. (a) Dependence on
      rotational state $j(\tau=0)$ for $j=0$ to $j=15$. (b) Dependence on the asymmetric-top quantum
      number $\tau$ for $j=0$ to $j=5$. Lines are drawn to guide the eye. See text for details.}
   \label{fig:capturestates}
\end{figure}
We also investigated the dependence of the capture rate constants on the rotational state of \AP. As
exemplified for the \cis conformer at the experimental collision energy in
\autoref{fig:capturestates}~(a), the capture rates slightly decrease as $j$ increases. Since the
rotational temperature is about 1.1~K, about $90~\%$ of the \AP population is confined to the
rotational states $j=1-5$ for which the relative difference of the rate constants is $<10$~\%.
\autoref{fig:capturestates}~(b) shows the dependence of the capture rate constant for the \cis
conformer on the asymmetric-top quantum number $\tau$. The dependence is only weak over the range of
states $j=0-5$.

By averaging over capture rate constants of all populated rotational states of \AP at 1.1~K and for
a collision energy of 0.123~eV, one obtains the effective capture rate constants
$k_{2,\cis}=2.7\times10^{-9}~\qcms$ and $k_{2,\trans}=1.8\times10^{-9}~\qcms$ and their ratio
$k_{2,\cis}/k_{2,\trans}=1.5$. These values are in good agreement with the experimentally observed
second-order rate constants for the reaction with \Ca in the excited $(4p)$ state, see
\autoref{sec:exp_result:detuning}.

\section{Summary and conclusions}
\label{sec:conclusions}
We have presented a new method for the characterization of conformer-specific chemical reactivities.
In a proof-of-concept study, the two conformers of \AP were spatially separated in a molecular beam
using the electrostatic deflector. Subsequently, the separated conformers reacted with a stationary
target of Coulomb-crystallized, laser-cooled \Ca ions. Second-order rate constants for the reactions
of the individual conformers with \Ca were obtained. The reaction rate for the \cis conformer was
found to be a factor of two larger than that for the \trans conformer. A detailed analysis of the
rate constants for the individual electronic states of \Ca showed that the observed reaction rates
are dominated by the reaction of \AP with electronically excited \Ca($4p$). The reaction rates of
\AP with \Ca in its $(4s)~^2S_{1/2}$ ground and $(3d)~^2D_{3/2}$ excited states were found to be two
to three orders of magnitude smaller. CaOH$^+$ and CaNH$_2^+$ were identified as the likely reaction
products by resonant-excitation mass spectrometry.

The rate constant observed for the \Ca($4p$) reaction channel was found to be close to the capture
limit. The difference in the reactivities of the two conformers could be rationalized in terms of
adiabatic-capture theory in very good agreement with the experimental findings. Within the capture
picture, the increased reaction rate for the \cis conformer compared to the \trans species is
explained by the stronger ion-dipole long-range interaction which results in a larger capture cross
section. The small reaction rates with \Ca in its $(4s)$ and $(3d)$ states indicate the existence
of dynamic bottlenecks along the reaction path. Preliminary DFT calculations for the reaction of \AP
with \Ca on the ground state potential energy surface enabled a first characterization of possible
reaction pathways. However, more extensive computations are necessary to elucidate the short-range
dynamics in all three reaction channels probed in the present experiments.

We expect that the present technique of combining electrostatic conformer selection with highly
sensitive Coulomb-crystal methods will enable the study of conformational effects in a range of
ion-molecule reactions. Electrostatic conformer separation is applicable to a variety of polar
molecules as long as their conformers exhibit appreciably different dipole moments. More advanced
techniques for the separation of molecular species and individual quantum states using electric
field manipulation have been reported and could be implemented in the current
methodology~\cite{Filsinger:PRL100:133003, Nielsen:PCCP13:18971, Putzke:PCCP13:18962,
   Trippel:PRA86:033202, Meerakker:CR112:2012}. For the ionic reaction partners, the generation of
Coulomb crystals of sympathetically cooled ions allows the study of a wide range of atomic and
molecular ionic species~\cite{Willitsch:IRPC31:175}. Moreover, the preparation of Coulomb crystals
with molecular ions in selected internal quantum states has been recently
accomplished~\cite{Tong:PRL105:143001, Tong:PRA83:023415} so that simultaneous studies of
conformational and state-specific effects are now within reach for a wide range of ion-molecule
reactions.

\begin{acknowledgments}
   This work has been supported by the Swiss National Science Foundation grant.\ nr.\
   PP00P2\_140834, the University of Basel, and the excellence cluster ``The Hamburg Center for
   Ultrafast Imaging -- Structure, Dynamics and Control of Matter at the Atomic Scale'' of the
   Deutsche Forschungsgemeinschaft.
\end{acknowledgments}


\begin{thebibliography}{75}%
\makeatletter
\providecommand \@ifxundefined [1]{%
 \@ifx{#1\undefined}
}%
\providecommand \@ifnum [1]{%
 \ifnum #1\expandafter \@firstoftwo
 \else \expandafter \@secondoftwo
 \fi
}%
\providecommand \@ifx [1]{%
 \ifx #1\expandafter \@firstoftwo
 \else \expandafter \@secondoftwo
 \fi
}%
\providecommand \natexlab [1]{#1}%
\providecommand \enquote  [1]{``#1''}%
\providecommand \bibnamefont  [1]{#1}%
\providecommand \bibfnamefont [1]{#1}%
\providecommand \citenamefont [1]{#1}%
\providecommand \href@noop [0]{\@secondoftwo}%
\providecommand \href [0]{\begingroup \@sanitize@url \@href}%
\providecommand \@href[1]{\@@startlink{#1}\@@href}%
\providecommand \@@href[1]{\endgroup#1\@@endlink}%
\providecommand \@sanitize@url [0]{\catcode `\\12\catcode `\$12\catcode
  `\&12\catcode `\#12\catcode `\^12\catcode `\_12\catcode `\%12\relax}%
\providecommand \@@startlink[1]{}%
\providecommand \@@endlink[0]{}%
\providecommand \url  [0]{\begingroup\@sanitize@url \@url }%
\providecommand \@url [1]{\endgroup\@href {#1}{\urlprefix }}%
\providecommand \urlprefix  [0]{URL }%
\providecommand \Eprint [0]{\href }%
\providecommand \doibase [0]{http://dx.doi.org/}%
\providecommand \selectlanguage [0]{\@gobble}%
\providecommand \bibinfo  [0]{\@secondoftwo}%
\providecommand \bibfield  [0]{\@secondoftwo}%
\providecommand \translation [1]{[#1]}%
\providecommand \BibitemOpen [0]{}%
\providecommand \bibitemStop [0]{}%
\providecommand \bibitemNoStop [0]{.\EOS\space}%
\providecommand \EOS [0]{\spacefactor3000\relax}%
\providecommand \BibitemShut  [1]{\csname bibitem#1\endcsname}%
\let\auto@bib@innerbib\@empty
\bibitem [{\citenamefont {Barton}(1953)}]{Barton:JCS1027}%
  \BibitemOpen
  \bibfield  {author} {\bibinfo {author} {\bibfnamefont {D.~H.~R.}\
  \bibnamefont {Barton}},\ }\bibfield  {title} {\enquote {\bibinfo {title} {The
  stereochemistry of cyclohexane derivatives},}\ }\href {\doibase
  10.1039/jr9530001027} {\bibfield  {journal} {\bibinfo  {journal} {J.\ Chem.\
  Soc.}\ }\textbf {\bibinfo {volume} {1953}},\ \bibinfo {pages} {1027}
  (\bibinfo {year} {1953})}\BibitemShut {NoStop}%
\bibitem [{\citenamefont {Dunathan}(1966)}]{Dunathan:PNAS55:712}%
  \BibitemOpen
  \bibfield  {author} {\bibinfo {author} {\bibfnamefont {H.~C.}\ \bibnamefont
  {Dunathan}},\ }\bibfield  {title} {\enquote {\bibinfo {title} {Conformation
  and reaction specificity in pyridoxal phosphate enzymes},}\ }\href {\doibase
  10.1073/pnas.55.4.712} {\bibfield  {journal} {\bibinfo  {journal} {PNAS}\
  }\textbf {\bibinfo {volume} {55}},\ \bibinfo {pages} {712--716} (\bibinfo
  {year} {1966})}\BibitemShut {NoStop}%
\bibitem [{\citenamefont {Eliel}\ and\ \citenamefont
  {Wilen}(1994)}]{Eliel:StereoChem:1994}%
  \BibitemOpen
  \bibfield  {author} {\bibinfo {author} {\bibfnamefont {E.~L.}\ \bibnamefont
  {Eliel}}\ and\ \bibinfo {author} {\bibfnamefont {S.~H.}\ \bibnamefont
  {Wilen}},\ }\href@noop {} {\emph {\bibinfo {title} {Stereochemistry of
  organic compounds}}}\ (\bibinfo  {publisher} {John Wiley \& Sons},\ \bibinfo
  {address} {New York},\ \bibinfo {year} {1994})\BibitemShut {NoStop}%
\bibitem [{\citenamefont {Suenram}\ and\ \citenamefont
  {Lovas}(1980)}]{Suenram:JACS102:7180}%
  \BibitemOpen
  \bibfield  {author} {\bibinfo {author} {\bibfnamefont {R.~D.}\ \bibnamefont
  {Suenram}}\ and\ \bibinfo {author} {\bibfnamefont {F.~J.}\ \bibnamefont
  {Lovas}},\ }\bibfield  {title} {\enquote {\bibinfo {title} {Millimeter wave
  spectrum of glycine - a new conformer},}\ }\href {\doibase
  10.1021/ja00544a002} {\bibfield  {journal} {\bibinfo  {journal} {J.\ Am.\
  Chem.\ Soc.}\ }\textbf {\bibinfo {volume} {102}},\ \bibinfo {pages}
  {7180--7184} (\bibinfo {year} {1980})}\BibitemShut {NoStop}%
\bibitem [{\citenamefont {Rizzo}\ \emph {et~al.}(1986)\citenamefont {Rizzo},
  \citenamefont {Park}, \citenamefont {Peteanu},\ and\ \citenamefont
  {Levy}}]{Rizzo:JCP84:2534}%
  \BibitemOpen
  \bibfield  {author} {\bibinfo {author} {\bibfnamefont {T.~R.}\ \bibnamefont
  {Rizzo}}, \bibinfo {author} {\bibfnamefont {Y.~D.}\ \bibnamefont {Park}},
  \bibinfo {author} {\bibfnamefont {L.~A.}\ \bibnamefont {Peteanu}}, \ and\
  \bibinfo {author} {\bibfnamefont {D.~H.}\ \bibnamefont {Levy}},\ }\bibfield
  {title} {\enquote {\bibinfo {title} {The electronic spectrum of the amino
  acid tryptophan in the gas phase},}\ }\href {\doibase 10.1063/1.450323}
  {\bibfield  {journal} {\bibinfo  {journal} {J.\ Chem.\ Phys.}\ }\textbf
  {\bibinfo {volume} {84}},\ \bibinfo {pages} {2534--2541} (\bibinfo {year}
  {1986})}\BibitemShut {NoStop}%
\bibitem [{\citenamefont {Robertson}\ and\ \citenamefont
  {Simons}(2001)}]{Robertson:PCCP3:1}%
  \BibitemOpen
  \bibfield  {author} {\bibinfo {author} {\bibfnamefont {E.~G.}\ \bibnamefont
  {Robertson}}\ and\ \bibinfo {author} {\bibfnamefont {J.~P.}\ \bibnamefont
  {Simons}},\ }\bibfield  {title} {\enquote {\bibinfo {title} {Getting into
  shape: Conformational and supramolecular landscapes in small biomolecules and
  their hydrated clusters},}\ }\href {\doibase 10.1039/B008225M} {\bibfield
  {journal} {\bibinfo  {journal} {Phys.\ Chem.\ Chem.\ Phys.}\ }\textbf
  {\bibinfo {volume} {3}},\ \bibinfo {pages} {1--18} (\bibinfo {year}
  {2001})}\BibitemShut {NoStop}%
\bibitem [{\citenamefont {Weinkauf}\ \emph {et~al.}(2002)\citenamefont
  {Weinkauf}, \citenamefont {Schermann}, \citenamefont {de~Vries},\ and\
  \citenamefont {Kleinermanns}}]{Weinkauf:EPJD20:309}%
  \BibitemOpen
  \bibfield  {author} {\bibinfo {author} {\bibfnamefont {R.}~\bibnamefont
  {Weinkauf}}, \bibinfo {author} {\bibfnamefont {J.}~\bibnamefont {Schermann}},
  \bibinfo {author} {\bibfnamefont {M.~S.}\ \bibnamefont {de~Vries}}, \ and\
  \bibinfo {author} {\bibfnamefont {K.}~\bibnamefont {Kleinermanns}},\
  }\bibfield  {title} {\enquote {\bibinfo {title} {Molecular physics of
  building blocks of life under isolated or defined conditions},}\ }\href
  {\doibase 10.1140/epjd/e2002-00185-0} {\bibfield  {journal} {\bibinfo
  {journal} {Eur.\ Phys.\ J.\ D}\ }\textbf {\bibinfo {volume} {20}},\ \bibinfo
  {pages} {309--316} (\bibinfo {year} {2002})}\BibitemShut {NoStop}%
\bibitem [{\citenamefont {Simons}(2004)}]{Simons:PCCP6:Biomolecules}%
  \BibitemOpen
  \bibfield  {author} {\bibinfo {author} {\bibfnamefont {J.~P.}\ \bibnamefont
  {Simons}},\ }\bibfield  {title} {\enquote {\bibinfo {title} {Bio-active
  molecules in the gas phase},}\ }\href {\doibase 10.1039/B405201N} {\bibfield
  {journal} {\bibinfo  {journal} {Phys.\ Chem.\ Chem.\ Phys.}\ }\textbf
  {\bibinfo {volume} {6}},\ \bibinfo {pages} {2543--2890} (\bibinfo {year}
  {2004})}\BibitemShut {NoStop}%
\bibitem [{\citenamefont {Simons}\ \emph {et~al.}(2005)\citenamefont {Simons},
  \citenamefont {Jockusch}, \citenamefont {\c{C}ar\c{c}abal}, \citenamefont
  {H\"{u}nig}, \citenamefont {Kroemer}, \citenamefont {Macleod},\ and\
  \citenamefont {Snoek}}]{Simons:IRPC24:489}%
  \BibitemOpen
  \bibfield  {author} {\bibinfo {author} {\bibfnamefont {J.~P.}\ \bibnamefont
  {Simons}}, \bibinfo {author} {\bibfnamefont {R.~A.}\ \bibnamefont
  {Jockusch}}, \bibinfo {author} {\bibfnamefont {P.}~\bibnamefont
  {\c{C}ar\c{c}abal}}, \bibinfo {author} {\bibfnamefont {I.}~\bibnamefont
  {H\"{u}nig}}, \bibinfo {author} {\bibfnamefont {R.~T.}\ \bibnamefont
  {Kroemer}}, \bibinfo {author} {\bibfnamefont {N.~A.}\ \bibnamefont
  {Macleod}}, \ and\ \bibinfo {author} {\bibfnamefont {L.~C.}\ \bibnamefont
  {Snoek}},\ }\bibfield  {title} {\enquote {\bibinfo {title} {Sugars in the gas
  phase. {S}pectroscopy, conformation, hydration, co-operativity and
  selectivity},}\ }\href {\doibase 10.1080/01442350500415107} {\bibfield
  {journal} {\bibinfo  {journal} {Int.\ Rev.\ Phys.\ Chem.}\ }\textbf {\bibinfo
  {volume} {24}},\ \bibinfo {pages} {489} (\bibinfo {year} {2005})}\BibitemShut
  {NoStop}%
\bibitem [{\citenamefont {de~Vries}\ and\ \citenamefont
  {Hobza}(2007)}]{Vries:ARPC58:585}%
  \BibitemOpen
  \bibfield  {author} {\bibinfo {author} {\bibfnamefont {M.~S.}\ \bibnamefont
  {de~Vries}}\ and\ \bibinfo {author} {\bibfnamefont {P.}~\bibnamefont
  {Hobza}},\ }\bibfield  {title} {\enquote {\bibinfo {title} {Gas-phase
  spectroscopy of biomolecular building blocks},}\ }\href {\doibase
  10.1146/annurev.physchem.57.032905.104722} {\bibfield  {journal} {\bibinfo
  {journal} {Annu.\ Rev.\ Phys.\ Chem.}\ }\textbf {\bibinfo {volume} {58}},\
  \bibinfo {pages} {585--612} (\bibinfo {year} {2007})}\BibitemShut {NoStop}%
\bibitem [{\citenamefont {Sanz}\ \emph {et~al.}(2008)\citenamefont {Sanz},
  \citenamefont {Blanco}, \citenamefont {Lopez},\ and\ \citenamefont
  {Alonso}}]{EugeniaSanz:ACIE47:6216}%
  \BibitemOpen
  \bibfield  {author} {\bibinfo {author} {\bibfnamefont {M.~E.}\ \bibnamefont
  {Sanz}}, \bibinfo {author} {\bibfnamefont {S.}~\bibnamefont {Blanco}},
  \bibinfo {author} {\bibfnamefont {J.~C.}\ \bibnamefont {Lopez}}, \ and\
  \bibinfo {author} {\bibfnamefont {J.~L.}\ \bibnamefont {Alonso}},\ }\bibfield
   {title} {\enquote {\bibinfo {title} {Rotational probes of six conformers of
  neutral cysteine},}\ }\href {\doibase 10.1002/anie.200801337} {\bibfield
  {journal} {\bibinfo  {journal} {Angew.\ Chem.\ Int.\ Ed.}\ }\textbf {\bibinfo
  {volume} {47}},\ \bibinfo {pages} {6216--6220} (\bibinfo {year}
  {2008})}\BibitemShut {NoStop}%
\bibitem [{\citenamefont {Rizzo}, \citenamefont {Stearns},\ and\ \citenamefont
  {Boyarkin}(2009)}]{Rizzo:IRPC28:481}%
  \BibitemOpen
  \bibfield  {author} {\bibinfo {author} {\bibfnamefont {T.~R.}\ \bibnamefont
  {Rizzo}}, \bibinfo {author} {\bibfnamefont {J.~A.}\ \bibnamefont {Stearns}},
  \ and\ \bibinfo {author} {\bibfnamefont {O.~V.}\ \bibnamefont {Boyarkin}},\
  }\bibfield  {title} {\enquote {\bibinfo {title} {Spectroscopic studies of
  cold, gas-phase biomolecular ions},}\ }\href {\doibase
  10.1080/01442350903069931} {\bibfield  {journal} {\bibinfo  {journal} {Int.\
  Rev.\ Phys.\ Chem.}\ }\textbf {\bibinfo {volume} {28}},\ \bibinfo {pages}
  {481} (\bibinfo {year} {2009})}\BibitemShut {NoStop}%
\bibitem [{\citenamefont {Nagornova}, \citenamefont {Rizzo},\ and\
  \citenamefont {Boyarkin}(2012)}]{Nagornova:Science336:320}%
  \BibitemOpen
  \bibfield  {author} {\bibinfo {author} {\bibfnamefont {N.~S.}\ \bibnamefont
  {Nagornova}}, \bibinfo {author} {\bibfnamefont {T.~R.}\ \bibnamefont
  {Rizzo}}, \ and\ \bibinfo {author} {\bibfnamefont {O.~V.}\ \bibnamefont
  {Boyarkin}},\ }\bibfield  {title} {\enquote {\bibinfo {title} {{Interplay of
  intra- and intermolecular H-bonding in a progressively solvated macrocyclic
  peptide.}}}\ }\href {\doibase 10.1126/science.1218709} {\bibfield  {journal}
  {\bibinfo  {journal} {Science}\ }\textbf {\bibinfo {volume} {336}},\ \bibinfo
  {pages} {320--3} (\bibinfo {year} {2012})}\BibitemShut {NoStop}%
\bibitem [{\citenamefont {Dian}, \citenamefont {Longarte},\ and\ \citenamefont
  {Zwier}(2002)}]{Dian:Science296:2369}%
  \BibitemOpen
  \bibfield  {author} {\bibinfo {author} {\bibfnamefont {B.~C.}\ \bibnamefont
  {Dian}}, \bibinfo {author} {\bibfnamefont {A.}~\bibnamefont {Longarte}}, \
  and\ \bibinfo {author} {\bibfnamefont {T.~S.}\ \bibnamefont {Zwier}},\
  }\bibfield  {title} {\enquote {\bibinfo {title} {Conformational dynamics in a
  dipeptide after single-mode vibrational excitation},}\ }\href {\doibase
  10.1126/science.1071563} {\bibfield  {journal} {\bibinfo  {journal}
  {Science}\ }\textbf {\bibinfo {volume} {296}},\ \bibinfo {pages} {2369--2373}
  (\bibinfo {year} {2002})}\BibitemShut {NoStop}%
\bibitem [{\citenamefont {Dian}, \citenamefont {Clarkson},\ and\ \citenamefont
  {Zwier}(2004)}]{Dian:Science303:1169}%
  \BibitemOpen
  \bibfield  {author} {\bibinfo {author} {\bibfnamefont {B.~C.}\ \bibnamefont
  {Dian}}, \bibinfo {author} {\bibfnamefont {J.~R.}\ \bibnamefont {Clarkson}},
  \ and\ \bibinfo {author} {\bibfnamefont {T.~S.}\ \bibnamefont {Zwier}},\
  }\bibfield  {title} {\enquote {\bibinfo {title} {Direct measurement of energy
  thresholds to conformational isomerization in tryptamine},}\ }\href {\doibase
  10.1126/science.1093731} {\bibfield  {journal} {\bibinfo  {journal}
  {Science}\ }\textbf {\bibinfo {volume} {303}},\ \bibinfo {pages} {1169--1173}
  (\bibinfo {year} {2004})}\BibitemShut {NoStop}%
\bibitem [{\citenamefont {Dian}\ \emph {et~al.}(2008)\citenamefont {Dian},
  \citenamefont {Brown}, \citenamefont {Douglass},\ and\ \citenamefont
  {Pate}}]{Dian:Science320:924}%
  \BibitemOpen
  \bibfield  {author} {\bibinfo {author} {\bibfnamefont {B.~C.}\ \bibnamefont
  {Dian}}, \bibinfo {author} {\bibfnamefont {G.~G.}\ \bibnamefont {Brown}},
  \bibinfo {author} {\bibfnamefont {K.~O.}\ \bibnamefont {Douglass}}, \ and\
  \bibinfo {author} {\bibfnamefont {B.~H.}\ \bibnamefont {Pate}},\ }\bibfield
  {title} {\enquote {\bibinfo {title} {Measuring picosecond isomerization
  kinetics via broadband microwave spectroscopy},}\ }\href {\doibase
  10.1126/science.1155736} {\bibfield  {journal} {\bibinfo  {journal}
  {Science}\ }\textbf {\bibinfo {volume} {320}},\ \bibinfo {pages} {924--928}
  (\bibinfo {year} {2008})}\BibitemShut {NoStop}%
\bibitem [{\citenamefont {Park}\ and\ \citenamefont
  {Kim}(2002)}]{Park:JACS124:7614}%
  \BibitemOpen
  \bibfield  {author} {\bibinfo {author} {\bibfnamefont {S.~T.}\ \bibnamefont
  {Park}}\ and\ \bibinfo {author} {\bibfnamefont {M.~S.}\ \bibnamefont {Kim}},\
  }\bibfield  {title} {\enquote {\bibinfo {title} {{Photodissociation dynamics
  of various conformers of iodobutane isomer ions prepared selectively by
  vacuum ultraviolet mass-analyzed threshold ionization.}}}\ }\href
  {http://pubs.acs.org/doi/abs/10.1021/ja025791e} {\bibfield  {journal}
  {\bibinfo  {journal} {J.\ Am.\ Chem.\ Soc.}\ }\textbf {\bibinfo {volume}
  {124}},\ \bibinfo {pages} {7614--21} (\bibinfo {year} {2002})}\BibitemShut
  {NoStop}%
\bibitem [{\citenamefont {Park}, \citenamefont {Kim},\ and\ \citenamefont
  {Kim}(2002)}]{Park:Nature415:306}%
  \BibitemOpen
  \bibfield  {author} {\bibinfo {author} {\bibfnamefont {S.~T.}\ \bibnamefont
  {Park}}, \bibinfo {author} {\bibfnamefont {S.~K.}\ \bibnamefont {Kim}}, \
  and\ \bibinfo {author} {\bibfnamefont {M.~S.}\ \bibnamefont {Kim}},\
  }\bibfield  {title} {\enquote {\bibinfo {title} {Observation of
  conformer-specific pathways in the photodissociation of 1-iodopropane
  ions},}\ }\href {http://dx.doi.org/10.1038/415306a} {\bibfield  {journal}
  {\bibinfo  {journal} {Nature}\ }\textbf {\bibinfo {volume} {415}},\ \bibinfo
  {pages} {306} (\bibinfo {year} {2002})}\BibitemShut {NoStop}%
\bibitem [{\citenamefont {Kim}\ \emph {et~al.}(2007)\citenamefont {Kim},
  \citenamefont {Shen}, \citenamefont {Tao}, \citenamefont {Martinez},\ and\
  \citenamefont {Suits}}]{Kim:Science315:1561}%
  \BibitemOpen
  \bibfield  {author} {\bibinfo {author} {\bibfnamefont {M.~H.}\ \bibnamefont
  {Kim}}, \bibinfo {author} {\bibfnamefont {L.}~\bibnamefont {Shen}}, \bibinfo
  {author} {\bibfnamefont {H.}~\bibnamefont {Tao}}, \bibinfo {author}
  {\bibfnamefont {T.~J.}\ \bibnamefont {Martinez}}, \ and\ \bibinfo {author}
  {\bibfnamefont {A.~G.}\ \bibnamefont {Suits}},\ }\bibfield  {title} {\enquote
  {\bibinfo {title} {Conformationally controlled chemistry: exited-state
  dynamics dictate ground-state reaction},}\ }\href {\doibase
  10.1126/science.1136453} {\bibfield  {journal} {\bibinfo  {journal}
  {Science}\ }\textbf {\bibinfo {volume} {315}},\ \bibinfo {pages} {1561}
  (\bibinfo {year} {2007})}\BibitemShut {NoStop}%
\bibitem [{\citenamefont {Oliver}, \citenamefont {King},\ and\ \citenamefont
  {Ashfold}(2010{\natexlab{a}})}]{Oliver:JCP133:194303}%
  \BibitemOpen
  \bibfield  {author} {\bibinfo {author} {\bibfnamefont {T.~A.~A.}\
  \bibnamefont {Oliver}}, \bibinfo {author} {\bibfnamefont {G.~A.}\
  \bibnamefont {King}}, \ and\ \bibinfo {author} {\bibfnamefont {M.~N.~R.}\
  \bibnamefont {Ashfold}},\ }\bibfield  {title} {\enquote {\bibinfo {title}
  {The ultraviolet photodissociation of axial and equatorial conformers of
  3-pyrroline.}}\ }\href {\doibase 10.1063/1.3503635} {\bibfield  {journal}
  {\bibinfo  {journal} {J.\ Chem.\ Phys.}\ }\textbf {\bibinfo {volume} {133}},\
  \bibinfo {pages} {194303} (\bibinfo {year} {2010}{\natexlab{a}})}\BibitemShut
  {NoStop}%
\bibitem [{\citenamefont {Oliver}, \citenamefont {King},\ and\ \citenamefont
  {Ashfold}(2010{\natexlab{b}})}]{Oliver:CS1:89}%
  \BibitemOpen
  \bibfield  {author} {\bibinfo {author} {\bibfnamefont {T.~A.~A.}\
  \bibnamefont {Oliver}}, \bibinfo {author} {\bibfnamefont {G.~A.}\
  \bibnamefont {King}}, \ and\ \bibinfo {author} {\bibfnamefont {M.~N.~R.}\
  \bibnamefont {Ashfold}},\ }\bibfield  {title} {\enquote {\bibinfo {title}
  {{The conformer resolved ultraviolet photodissociation of morpholine}},}\
  }\href {\doibase 10.1039/c0sc00119h} {\bibfield  {journal} {\bibinfo
  {journal} {Chem.\ Sci.}\ }\textbf {\bibinfo {volume} {1}},\ \bibinfo {pages}
  {89} (\bibinfo {year} {2010}{\natexlab{b}})}\BibitemShut {NoStop}%
\bibitem [{\citenamefont {Zaouris}\ \emph {et~al.}(2011)\citenamefont
  {Zaouris}, \citenamefont {Wenge}, \citenamefont {Murdock}, \citenamefont
  {Oliver}, \citenamefont {Richmond}, \citenamefont {Ritchie}, \citenamefont
  {Dixon},\ and\ \citenamefont {Ashfold}}]{Zaouris:JCP135:094312}%
  \BibitemOpen
  \bibfield  {author} {\bibinfo {author} {\bibfnamefont {D.~K.}\ \bibnamefont
  {Zaouris}}, \bibinfo {author} {\bibfnamefont {A.~M.}\ \bibnamefont {Wenge}},
  \bibinfo {author} {\bibfnamefont {D.}~\bibnamefont {Murdock}}, \bibinfo
  {author} {\bibfnamefont {T.~A.~A.}\ \bibnamefont {Oliver}}, \bibinfo {author}
  {\bibfnamefont {G.}~\bibnamefont {Richmond}}, \bibinfo {author}
  {\bibfnamefont {G.~A.~D.}\ \bibnamefont {Ritchie}}, \bibinfo {author}
  {\bibfnamefont {R.~N.}\ \bibnamefont {Dixon}}, \ and\ \bibinfo {author}
  {\bibfnamefont {M.~N.~R.}\ \bibnamefont {Ashfold}},\ }\bibfield  {title}
  {\enquote {\bibinfo {title} {{Conformer specific dissociation dynamics of
  iodocyclohexane studied by velocity map imaging.}}}\ }\href {\doibase
  10.1063/1.3628682} {\bibfield  {journal} {\bibinfo  {journal} {J.\ Chem.\
  Phys.}\ }\textbf {\bibinfo {volume} {135}},\ \bibinfo {pages} {094312}
  (\bibinfo {year} {2011})}\BibitemShut {NoStop}%
\bibitem [{\citenamefont {Taatjes}\ \emph {et~al.}(2013)\citenamefont
  {Taatjes}, \citenamefont {Welz}, \citenamefont {Eskola}, \citenamefont
  {Savee}, \citenamefont {Scheer}, \citenamefont {Shallcross}, \citenamefont
  {Rotavera}, \citenamefont {Lee}, \citenamefont {Dyke}, \citenamefont {Mok},
  \citenamefont {Osborn},\ and\ \citenamefont
  {Percival}}]{Taatjes:Science340:177}%
  \BibitemOpen
  \bibfield  {author} {\bibinfo {author} {\bibfnamefont {C.~A.}\ \bibnamefont
  {Taatjes}}, \bibinfo {author} {\bibfnamefont {O.}~\bibnamefont {Welz}},
  \bibinfo {author} {\bibfnamefont {A.~J.}\ \bibnamefont {Eskola}}, \bibinfo
  {author} {\bibfnamefont {J.~D.}\ \bibnamefont {Savee}}, \bibinfo {author}
  {\bibfnamefont {A.~M.}\ \bibnamefont {Scheer}}, \bibinfo {author}
  {\bibfnamefont {D.~E.}\ \bibnamefont {Shallcross}}, \bibinfo {author}
  {\bibfnamefont {B.}~\bibnamefont {Rotavera}}, \bibinfo {author}
  {\bibfnamefont {E.~P.~F.}\ \bibnamefont {Lee}}, \bibinfo {author}
  {\bibfnamefont {J.~M.}\ \bibnamefont {Dyke}}, \bibinfo {author}
  {\bibfnamefont {D.~K.~W.}\ \bibnamefont {Mok}}, \bibinfo {author}
  {\bibfnamefont {D.~L.}\ \bibnamefont {Osborn}}, \ and\ \bibinfo {author}
  {\bibfnamefont {C.~J.}\ \bibnamefont {Percival}},\ }\bibfield  {title}
  {\enquote {\bibinfo {title} {Direct measurement of conformer-dependent
  reactivity of the criegee intermediate {CH3CHOO}},}\ }\href {\doibase
  10.1126/science.1234689} {\bibfield  {journal} {\bibinfo  {journal}
  {Science}\ }\textbf {\bibinfo {volume} {340}},\ \bibinfo {pages} {177}
  (\bibinfo {year} {2013})}\BibitemShut {NoStop}%
\bibitem [{\citenamefont {Khriachtchev}\ \emph {et~al.}(2009)\citenamefont
  {Khriachtchev}, \citenamefont {Domanskaya}, \citenamefont {Marushkevich},
  \citenamefont {R\"{a}s\"{a}nen}, \citenamefont {Grigorenko}, \citenamefont
  {Ermilov}, \citenamefont {Andrijchenko},\ and\ \citenamefont
  {Nemukhin}}]{Khriachtchev:JPCA113:8143}%
  \BibitemOpen
  \bibfield  {author} {\bibinfo {author} {\bibfnamefont {L.}~\bibnamefont
  {Khriachtchev}}, \bibinfo {author} {\bibfnamefont {A.}~\bibnamefont
  {Domanskaya}}, \bibinfo {author} {\bibfnamefont {K.}~\bibnamefont
  {Marushkevich}}, \bibinfo {author} {\bibfnamefont {M.}~\bibnamefont
  {R\"{a}s\"{a}nen}}, \bibinfo {author} {\bibfnamefont {B.}~\bibnamefont
  {Grigorenko}}, \bibinfo {author} {\bibfnamefont {A.}~\bibnamefont {Ermilov}},
  \bibinfo {author} {\bibfnamefont {N.}~\bibnamefont {Andrijchenko}}, \ and\
  \bibinfo {author} {\bibfnamefont {A.}~\bibnamefont {Nemukhin}},\ }\bibfield
  {title} {\enquote {\bibinfo {title} {Conformation-dependent chemical reaction
  of formic acid with an oxygen atom},}\ }\href {\doibase 10.1021/jp903775k}
  {\bibfield  {journal} {\bibinfo  {journal} {J.\ Phys.\ Chem.\ A}\ }\textbf
  {\bibinfo {volume} {113}},\ \bibinfo {pages} {8143} (\bibinfo {year}
  {2009})}\BibitemShut {NoStop}%
\bibitem [{\citenamefont {von Helden}, \citenamefont {Wyttenbach},\ and\
  \citenamefont {Bowers}(1995)}]{Helden:Science267:1483}%
  \BibitemOpen
  \bibfield  {author} {\bibinfo {author} {\bibfnamefont {G.}~\bibnamefont {von
  Helden}}, \bibinfo {author} {\bibfnamefont {T.}~\bibnamefont {Wyttenbach}}, \
  and\ \bibinfo {author} {\bibfnamefont {M.~T.}\ \bibnamefont {Bowers}},\
  }\bibfield  {title} {\enquote {\bibinfo {title} {Conformation of
  macromolecules in the gas-phase -- use of matrix-assisted laser-desorption
  methods in ion chromatography},}\ }\href {\doibase
  10.1126/science.267.5203.1483} {\bibfield  {journal} {\bibinfo  {journal}
  {Science}\ }\textbf {\bibinfo {volume} {267}},\ \bibinfo {pages} {1483--1485}
  (\bibinfo {year} {1995})}\BibitemShut {NoStop}%
\bibitem [{\citenamefont {Filsinger}\ \emph
  {et~al.}(2008{\natexlab{a}})\citenamefont {Filsinger}, \citenamefont
  {Erlekam}, \citenamefont {von Helden}, \citenamefont {K\"upper},\ and\
  \citenamefont {Meijer}}]{Filsinger:PRL100:133003}%
  \BibitemOpen
  \bibfield  {author} {\bibinfo {author} {\bibfnamefont {F.}~\bibnamefont
  {Filsinger}}, \bibinfo {author} {\bibfnamefont {U.}~\bibnamefont {Erlekam}},
  \bibinfo {author} {\bibfnamefont {G.}~\bibnamefont {von Helden}}, \bibinfo
  {author} {\bibfnamefont {J.}~\bibnamefont {K\"upper}}, \ and\ \bibinfo
  {author} {\bibfnamefont {G.}~\bibnamefont {Meijer}},\ }\bibfield  {title}
  {\enquote {\bibinfo {title} {Selector for structural isomers of neutral
  molecules},}\ }\href {\doibase 10.1103/PhysRevLett.100.133003} {\bibfield
  {journal} {\bibinfo  {journal} {Phys.\ Rev.\ Lett.}\ }\textbf {\bibinfo
  {volume} {100}},\ \bibinfo {pages} {133003} (\bibinfo {year}
  {2008}{\natexlab{a}})},\ \Eprint {http://arxiv.org/abs/0802.2795}
  {arXiv:0802.2795 [physics]} \BibitemShut {NoStop}%
\bibitem [{\citenamefont {Filsinger}\ \emph
  {et~al.}(2009{\natexlab{a}})\citenamefont {Filsinger}, \citenamefont
  {K\"upper}, \citenamefont {Meijer}, \citenamefont {Hansen}, \citenamefont
  {Maurer}, \citenamefont {Nielsen}, \citenamefont {Holmegaard},\ and\
  \citenamefont {Stapelfeldt}}]{Filsinger:ACIE48:6900}%
  \BibitemOpen
  \bibfield  {author} {\bibinfo {author} {\bibfnamefont {F.}~\bibnamefont
  {Filsinger}}, \bibinfo {author} {\bibfnamefont {J.}~\bibnamefont {K\"upper}},
  \bibinfo {author} {\bibfnamefont {G.}~\bibnamefont {Meijer}}, \bibinfo
  {author} {\bibfnamefont {J.~L.}\ \bibnamefont {Hansen}}, \bibinfo {author}
  {\bibfnamefont {J.}~\bibnamefont {Maurer}}, \bibinfo {author} {\bibfnamefont
  {J.~H.}\ \bibnamefont {Nielsen}}, \bibinfo {author} {\bibfnamefont
  {L.}~\bibnamefont {Holmegaard}}, \ and\ \bibinfo {author} {\bibfnamefont
  {H.}~\bibnamefont {Stapelfeldt}},\ }\bibfield  {title} {\enquote {\bibinfo
  {title} {Pure samples of individual conformers: the separation of
  stereo-isomers of complex molecules using electric fields},}\ }\href
  {\doibase 10.1002/anie.200902650} {\bibfield  {journal} {\bibinfo  {journal}
  {Angew.\ Chem.\ Int.\ Ed.}\ }\textbf {\bibinfo {volume} {48}},\ \bibinfo
  {pages} {6900--6902} (\bibinfo {year} {2009}{\natexlab{a}})}\BibitemShut
  {NoStop}%
\bibitem [{\citenamefont {Kierspel}\ \emph {et~al.}(2014)\citenamefont
  {Kierspel}, \citenamefont {Horke}, \citenamefont {Chang},\ and\ \citenamefont
  {Küpper}}]{Kierspel:CPL591:130}%
  \BibitemOpen
  \bibfield  {author} {\bibinfo {author} {\bibfnamefont {T.}~\bibnamefont
  {Kierspel}}, \bibinfo {author} {\bibfnamefont {D.~A.}\ \bibnamefont {Horke}},
  \bibinfo {author} {\bibfnamefont {Y.-P.}\ \bibnamefont {Chang}}, \ and\
  \bibinfo {author} {\bibfnamefont {J.}~\bibnamefont {Küpper}},\ }\bibfield
  {title} {\enquote {\bibinfo {title} {Spatially separated polar samples of the
  cis and trans conformers of 3-fluorophenol},}\ }\href {\doibase
  http://dx.doi.org/10.1016/j.cplett.2013.11.010} {\bibfield  {journal}
  {\bibinfo  {journal} {Chemical Physics Letters}\ }\textbf {\bibinfo {volume}
  {591}},\ \bibinfo {pages} {130--132} (\bibinfo {year} {2014})},\ \Eprint
  {http://arxiv.org/abs/1312.4417} {arXiv:1312.4417 [physics]} \BibitemShut
  {NoStop}%
\bibitem [{\citenamefont {Chang}\ \emph {et~al.}(2013)\citenamefont {Chang},
  \citenamefont {D{\l}ugo\l\k{e}cki}, \citenamefont {K\"upper}, \citenamefont
  {R\"osch}, \citenamefont {Wild},\ and\ \citenamefont
  {Willitsch}}]{Chang:Science342:98}%
  \BibitemOpen
  \bibfield  {author} {\bibinfo {author} {\bibfnamefont {Y.-P.}\ \bibnamefont
  {Chang}}, \bibinfo {author} {\bibfnamefont {K.}~\bibnamefont
  {D{\l}ugo\l\k{e}cki}}, \bibinfo {author} {\bibfnamefont {J.}~\bibnamefont
  {K\"upper}}, \bibinfo {author} {\bibfnamefont {D.}~\bibnamefont {R\"osch}},
  \bibinfo {author} {\bibfnamefont {D.}~\bibnamefont {Wild}}, \ and\ \bibinfo
  {author} {\bibfnamefont {S.}~\bibnamefont {Willitsch}},\ }\bibfield  {title}
  {\enquote {\bibinfo {title} {Specific chemical reactivities of spatially
  separated 3-aminophenol conformers with cold {Ca$^+$} ions},}\ }\href
  {\doibase 10.1126/science.1242271} {\bibfield  {journal} {\bibinfo  {journal}
  {Science}\ }\textbf {\bibinfo {volume} {342}},\ \bibinfo {pages} {98--101}
  (\bibinfo {year} {2013})},\ \Eprint {http://arxiv.org/abs/1308.6538}
  {arXiv:1308.6538 [physics]} \BibitemShut {NoStop}%
\bibitem [{\citenamefont {Willitsch}(2012)}]{Willitsch:IRPC31:175}%
  \BibitemOpen
  \bibfield  {author} {\bibinfo {author} {\bibfnamefont {S.}~\bibnamefont
  {Willitsch}},\ }\href {\doibase 10.1080/0144235X.2012.667221} {\bibfield
  {journal} {\bibinfo  {journal} {Int.\ Rev.\ Phys.\ Chem.}\ }\textbf {\bibinfo
  {volume} {31}},\ \bibinfo {pages} {175} (\bibinfo {year} {2012})}\BibitemShut
  {NoStop}%
\bibitem [{\citenamefont {Eller}\ and\ \citenamefont
  {Schwarz}(1991)}]{Eller:CR91:1121}%
  \BibitemOpen
  \bibfield  {author} {\bibinfo {author} {\bibfnamefont {K.}~\bibnamefont
  {Eller}}\ and\ \bibinfo {author} {\bibfnamefont {H.}~\bibnamefont
  {Schwarz}},\ }\bibfield  {title} {\enquote {\bibinfo {title} {Organometallic
  chemistry in the gas phase},}\ }\href {\doibase 10.1021/cr00006a002}
  {\bibfield  {journal} {\bibinfo  {journal} {Chem.\ Rev.}\ }\textbf {\bibinfo
  {volume} {91}},\ \bibinfo {pages} {1121--1177} (\bibinfo {year}
  {1991})}\BibitemShut {NoStop}%
\bibitem [{\citenamefont {Schwarz}(2011)}]{Schwarz:ACIE50:10096}%
  \BibitemOpen
  \bibfield  {author} {\bibinfo {author} {\bibfnamefont {H.}~\bibnamefont
  {Schwarz}},\ }\bibfield  {title} {\enquote {\bibinfo {title} {Chemistry with
  methane: Concepts rather than recipes},}\ }\href {\doibase
  10.1002/anie.201006424} {\bibfield  {journal} {\bibinfo  {journal} {Angew.\
  Chem.\ Int.\ Ed.}\ }\textbf {\bibinfo {volume} {50}},\ \bibinfo {pages}
  {10096} (\bibinfo {year} {2011})}\BibitemShut {NoStop}%
\bibitem [{\citenamefont {Harvey}\ \emph {et~al.}(1997)\citenamefont {Harvey},
  \citenamefont {Schr\"{o}der}, \citenamefont {Koch}, \citenamefont {Danovich},
  \citenamefont {Shaik},\ and\ \citenamefont {Schwarz}}]{Harvey:CPL273:164}%
  \BibitemOpen
  \bibfield  {author} {\bibinfo {author} {\bibfnamefont {J.~N.}\ \bibnamefont
  {Harvey}}, \bibinfo {author} {\bibfnamefont {D.}~\bibnamefont
  {Schr\"{o}der}}, \bibinfo {author} {\bibfnamefont {W.}~\bibnamefont {Koch}},
  \bibinfo {author} {\bibfnamefont {D.}~\bibnamefont {Danovich}}, \bibinfo
  {author} {\bibfnamefont {S.}~\bibnamefont {Shaik}}, \ and\ \bibinfo {author}
  {\bibfnamefont {H.}~\bibnamefont {Schwarz}},\ }\href
  {http://dx.doi.org/10.1016/S0009-2614(97)01112-3} {\bibfield  {journal}
  {\bibinfo  {journal} {Chem.\ Phys.\ Lett.}\ }\textbf {\bibinfo {volume}
  {273}},\ \bibinfo {pages} {164} (\bibinfo {year} {1997})}\BibitemShut
  {NoStop}%
\bibitem [{\citenamefont {Zhao}, \citenamefont {Koyanagi},\ and\ \citenamefont
  {Bohme}(2006)}]{Zhao:JPCA110:10607}%
  \BibitemOpen
  \bibfield  {author} {\bibinfo {author} {\bibfnamefont {X.}~\bibnamefont
  {Zhao}}, \bibinfo {author} {\bibfnamefont {G.~K.}\ \bibnamefont {Koyanagi}},
  \ and\ \bibinfo {author} {\bibfnamefont {D.~K.}\ \bibnamefont {Bohme}},\
  }\bibfield  {title} {\enquote {\bibinfo {title} {Reactions of methyl fluoride
  with atomic transition-metal and main-group cations: Gas-phase
  room-temperature kinetics and periodicities in reactivity},}\ }\href
  {\doibase 10.1021/jp062625a} {\bibfield  {journal} {\bibinfo  {journal} {J.\
  Phys.\ Chem.\ A}\ }\textbf {\bibinfo {volume} {110}},\ \bibinfo {pages}
  {10607} (\bibinfo {year} {2006})}\BibitemShut {NoStop}%
\bibitem [{\citenamefont {Ryzhov}\ and\ \citenamefont
  {Dunbar}(1999)}]{Ryzhov:JACS121:2259}%
  \BibitemOpen
  \bibfield  {author} {\bibinfo {author} {\bibfnamefont {V.}~\bibnamefont
  {Ryzhov}}\ and\ \bibinfo {author} {\bibfnamefont {R.~C.}\ \bibnamefont
  {Dunbar}},\ }\bibfield  {title} {\enquote {\bibinfo {title} {Interactions of
  phenol and indole with metal ions in the gas phase: Models for {T}yr and
  {T}rp side-chain binding},}\ }\href {\doibase 10.1021/ja983272z} {\bibfield
  {journal} {\bibinfo  {journal} {J.\ Am.\ Chem.\ Soc.}\ }\textbf {\bibinfo
  {volume} {121}},\ \bibinfo {pages} {2259--2268} (\bibinfo {year}
  {1999})}\BibitemShut {NoStop}%
\bibitem [{\citenamefont {Filsinger}\ \emph
  {et~al.}(2009{\natexlab{b}})\citenamefont {Filsinger}, \citenamefont
  {K\"upper}, \citenamefont {Meijer}, \citenamefont {Holmegaard}, \citenamefont
  {Nielsen}, \citenamefont {Nevo}, \citenamefont {Hansen},\ and\ \citenamefont
  {Stapelfeldt}}]{Filsinger:JCP131:064309}%
  \BibitemOpen
  \bibfield  {author} {\bibinfo {author} {\bibfnamefont {F.}~\bibnamefont
  {Filsinger}}, \bibinfo {author} {\bibfnamefont {J.}~\bibnamefont {K\"upper}},
  \bibinfo {author} {\bibfnamefont {G.}~\bibnamefont {Meijer}}, \bibinfo
  {author} {\bibfnamefont {L.}~\bibnamefont {Holmegaard}}, \bibinfo {author}
  {\bibfnamefont {J.~H.}\ \bibnamefont {Nielsen}}, \bibinfo {author}
  {\bibfnamefont {I.}~\bibnamefont {Nevo}}, \bibinfo {author} {\bibfnamefont
  {J.~L.}\ \bibnamefont {Hansen}}, \ and\ \bibinfo {author} {\bibfnamefont
  {H.}~\bibnamefont {Stapelfeldt}},\ }\bibfield  {title} {\enquote {\bibinfo
  {title} {Quantum-state selection, alignment, and orientation of large
  molecules using static electric and laser fields},}\ }\href {\doibase
  10.1063/1.3194287} {\bibfield  {journal} {\bibinfo  {journal} {J.\ Chem.\
  Phys.}\ }\textbf {\bibinfo {volume} {131}},\ \bibinfo {pages} {064309}
  (\bibinfo {year} {2009}{\natexlab{b}})},\ \Eprint
  {http://arxiv.org/abs/0903.5413} {arXiv:0903.5413 [physics]} \BibitemShut
  {NoStop}%
\bibitem [{\citenamefont {Willitsch}\ \emph
  {et~al.}(2008{\natexlab{a}})\citenamefont {Willitsch}, \citenamefont {Bell},
  \citenamefont {Gingell},\ and\ \citenamefont
  {Softley}}]{Willitsch:PCCP10:7200}%
  \BibitemOpen
  \bibfield  {author} {\bibinfo {author} {\bibfnamefont {S.}~\bibnamefont
  {Willitsch}}, \bibinfo {author} {\bibfnamefont {M.~T.}\ \bibnamefont {Bell}},
  \bibinfo {author} {\bibfnamefont {A.~D.}\ \bibnamefont {Gingell}}, \ and\
  \bibinfo {author} {\bibfnamefont {T.~P.}\ \bibnamefont {Softley}},\
  }\bibfield  {title} {\enquote {\bibinfo {title} {Chemical applications of
  laser- and sympathetically-cooled ions in ion traps},}\ }\href
  {http://dx.doi.org/10.1039/b813408c} {\bibfield  {journal} {\bibinfo
  {journal} {Phys.\ Chem.\ Chem.\ Phys.}\ }\textbf {\bibinfo {volume} {10}},\
  \bibinfo {pages} {7200} (\bibinfo {year} {2008}{\natexlab{a}})}\BibitemShut
  {NoStop}%
\bibitem [{\citenamefont {Even}\ \emph {et~al.}(2000)\citenamefont {Even},
  \citenamefont {Jortner}, \citenamefont {Noy}, \citenamefont {Lavie},\ and\
  \citenamefont {Cossart-Magos}}]{Even:JCP112:8068}%
  \BibitemOpen
  \bibfield  {author} {\bibinfo {author} {\bibfnamefont {U.}~\bibnamefont
  {Even}}, \bibinfo {author} {\bibfnamefont {J.}~\bibnamefont {Jortner}},
  \bibinfo {author} {\bibfnamefont {D.}~\bibnamefont {Noy}}, \bibinfo {author}
  {\bibfnamefont {N.}~\bibnamefont {Lavie}}, \ and\ \bibinfo {author}
  {\bibfnamefont {N.}~\bibnamefont {Cossart-Magos}},\ }\bibfield  {title}
  {\enquote {\bibinfo {title} {Cooling of large molecules below 1~{K} and {H}e
  clusters formation},}\ }\href {\doibase 10.1063/1.481405} {\bibfield
  {journal} {\bibinfo  {journal} {J.\ Chem.\ Phys.}\ }\textbf {\bibinfo
  {volume} {112}},\ \bibinfo {pages} {8068--8071} (\bibinfo {year}
  {2000})}\BibitemShut {NoStop}%
\bibitem [{\citenamefont {Hughes}(1947)}]{Hughes:PR72:614}%
  \BibitemOpen
  \bibfield  {author} {\bibinfo {author} {\bibfnamefont {H.~K.}\ \bibnamefont
  {Hughes}},\ }\bibfield  {title} {\enquote {\bibinfo {title} {The electric
  resonance method of radiofrequency spectroscopy the moment of inertia and
  electric dipole moment of {CsF}},}\ }\href {\doibase 10.1103/PhysRev.72.614}
  {\bibfield  {journal} {\bibinfo  {journal} {Phys.\ Rev.}\ }\textbf {\bibinfo
  {volume} {72}},\ \bibinfo {pages} {614--623} (\bibinfo {year}
  {1947})}\BibitemShut {NoStop}%
\bibitem [{\citenamefont {Lee}\ \emph {et~al.}(1953)\citenamefont {Lee},
  \citenamefont {Fabricand}, \citenamefont {Carlson},\ and\ \citenamefont
  {Rabi}}]{Lee:PR91:1395}%
  \BibitemOpen
  \bibfield  {author} {\bibinfo {author} {\bibfnamefont {C.~A.}\ \bibnamefont
  {Lee}}, \bibinfo {author} {\bibfnamefont {B.~P.}\ \bibnamefont {Fabricand}},
  \bibinfo {author} {\bibfnamefont {R.~O.}\ \bibnamefont {Carlson}}, \ and\
  \bibinfo {author} {\bibfnamefont {I.~I.}\ \bibnamefont {Rabi}},\ }\bibfield
  {title} {\enquote {\bibinfo {title} {Molecular beam investigation of
  rotational transitions .1. the rotational levels of {KC}l and their hyperfine
  structure},}\ }\href@noop {} {\bibfield  {journal} {\bibinfo  {journal}
  {Phys.\ Rev.}\ }\textbf {\bibinfo {volume} {91}},\ \bibinfo {pages}
  {1395--1403} (\bibinfo {year} {1953})}\BibitemShut {NoStop}%
\bibitem [{\citenamefont {Ramsey}(1956)}]{Ramsey:MolBeam:1956}%
  \BibitemOpen
  \bibfield  {author} {\bibinfo {author} {\bibfnamefont {N.~F.}\ \bibnamefont
  {Ramsey}},\ }\href@noop {} {\emph {\bibinfo {title} {Molecular Beams}}},\ The
  International Series of Monographs on Physics\ (\bibinfo  {publisher} {Oxford
  University Press},\ \bibinfo {address} {London, GB},\ \bibinfo {year}
  {1956})\ \bibinfo {note} {reprinted in \emph{Oxford Classic Texts in the
  Physical Sciences} (2005)}\BibitemShut {NoStop}%
\bibitem [{\citenamefont {Holmegaard}\ \emph {et~al.}(2009)\citenamefont
  {Holmegaard}, \citenamefont {Nielsen}, \citenamefont {Nevo}, \citenamefont
  {Stapelfeldt}, \citenamefont {Filsinger}, \citenamefont {K\"upper},\ and\
  \citenamefont {Meijer}}]{Holmegaard:PRL102:023001}%
  \BibitemOpen
  \bibfield  {author} {\bibinfo {author} {\bibfnamefont {L.}~\bibnamefont
  {Holmegaard}}, \bibinfo {author} {\bibfnamefont {J.~H.}\ \bibnamefont
  {Nielsen}}, \bibinfo {author} {\bibfnamefont {I.}~\bibnamefont {Nevo}},
  \bibinfo {author} {\bibfnamefont {H.}~\bibnamefont {Stapelfeldt}}, \bibinfo
  {author} {\bibfnamefont {F.}~\bibnamefont {Filsinger}}, \bibinfo {author}
  {\bibfnamefont {J.}~\bibnamefont {K\"upper}}, \ and\ \bibinfo {author}
  {\bibfnamefont {G.}~\bibnamefont {Meijer}},\ }\bibfield  {title} {\enquote
  {\bibinfo {title} {Laser-induced alignment and orientation of
  quantum-state-selected large molecules},}\ }\href {\doibase
  10.1103/PhysRevLett.102.023001} {\bibfield  {journal} {\bibinfo  {journal}
  {Phys.\ Rev.\ Lett.}\ }\textbf {\bibinfo {volume} {102}},\ \bibinfo {pages}
  {023001} (\bibinfo {year} {2009})},\ \Eprint {http://arxiv.org/abs/0810.2307}
  {arXiv:0810.2307 [physics]} \BibitemShut {NoStop}%
\bibitem [{\citenamefont {Hall}\ \emph {et~al.}(2011)\citenamefont {Hall},
  \citenamefont {Aymar}, \citenamefont {Bouloufa-Maafa}, \citenamefont
  {Dulieu},\ and\ \citenamefont {Willitsch}}]{Hall:PRL107:2432002}%
  \BibitemOpen
  \bibfield  {author} {\bibinfo {author} {\bibfnamefont {F.~H.~J.}\
  \bibnamefont {Hall}}, \bibinfo {author} {\bibfnamefont {M.}~\bibnamefont
  {Aymar}}, \bibinfo {author} {\bibfnamefont {N.}~\bibnamefont
  {Bouloufa-Maafa}}, \bibinfo {author} {\bibfnamefont {O.}~\bibnamefont
  {Dulieu}}, \ and\ \bibinfo {author} {\bibfnamefont {S.}~\bibnamefont
  {Willitsch}},\ }\bibfield  {title} {\enquote {\bibinfo {title}
  {Light-assisted ion-neutral reactive processes in the cold regime: Radiative
  molecule formation versus charge exchange},}\ }\href {\doibase
  10.1103/PhysRevLett.107.243202} {\bibfield  {journal} {\bibinfo  {journal}
  {Phys.\ Rev.\ Lett.}\ }\textbf {\bibinfo {volume} {107}},\ \bibinfo {pages}
  {243202} (\bibinfo {year} {2011})},\ \Eprint {http://arxiv.org/abs/1301.0724}
  {arXiv:1301.0724 [physics]} \BibitemShut {NoStop}%
\bibitem [{\citenamefont {Willitsch}\ \emph
  {et~al.}(2008{\natexlab{b}})\citenamefont {Willitsch}, \citenamefont {Bell},
  \citenamefont {Gingell}, \citenamefont {Procter},\ and\ \citenamefont
  {Softley}}]{Willitsch:PRL100:043203}%
  \BibitemOpen
  \bibfield  {author} {\bibinfo {author} {\bibfnamefont {S.}~\bibnamefont
  {Willitsch}}, \bibinfo {author} {\bibfnamefont {M.~T.}\ \bibnamefont {Bell}},
  \bibinfo {author} {\bibfnamefont {A.~D.}\ \bibnamefont {Gingell}}, \bibinfo
  {author} {\bibfnamefont {S.~R.}\ \bibnamefont {Procter}}, \ and\ \bibinfo
  {author} {\bibfnamefont {T.~P.}\ \bibnamefont {Softley}},\ }\bibfield
  {title} {\enquote {\bibinfo {title} {Cold reactive collisions between
  laser-cooled ions and velocity-selected neutral molecules},}\ }\href
  {\doibase 10.1103/PhysRevLett.100.043203} {\bibfield  {journal} {\bibinfo
  {journal} {Phys.\ Rev.\ Lett.}\ }\textbf {\bibinfo {volume} {100}},\ \bibinfo
  {eid} {043203} (\bibinfo {year} {2008}{\natexlab{b}})}\BibitemShut {NoStop}%
\bibitem [{\citenamefont {Hall}\ \emph {et~al.}(2013)\citenamefont {Hall},
  \citenamefont {Eberle}, \citenamefont {Hegi}, \citenamefont {Raoult},
  \citenamefont {Aymar}, \citenamefont {Dulieu},\ and\ \citenamefont
  {Willitsch}}]{Hall:MP111:2020}%
  \BibitemOpen
  \bibfield  {author} {\bibinfo {author} {\bibfnamefont {F.~H.}\ \bibnamefont
  {Hall}}, \bibinfo {author} {\bibfnamefont {P.}~\bibnamefont {Eberle}},
  \bibinfo {author} {\bibfnamefont {G.}~\bibnamefont {Hegi}}, \bibinfo {author}
  {\bibfnamefont {M.}~\bibnamefont {Raoult}}, \bibinfo {author} {\bibfnamefont
  {M.}~\bibnamefont {Aymar}}, \bibinfo {author} {\bibfnamefont
  {O.}~\bibnamefont {Dulieu}}, \ and\ \bibinfo {author} {\bibfnamefont
  {S.}~\bibnamefont {Willitsch}},\ }\bibfield  {title} {\enquote {\bibinfo
  {title} {Ion-neutral chemistry at ultralow energies: dynamics of reactive
  collisions between laser-cooled {Ca$^+$} ions and {Rb} atoms in an ion-atom
  hybrid trap},}\ }\href {\doibase 10.1080/00268976.2013.780107} {\bibfield
  {journal} {\bibinfo  {journal} {Mol.\ Phys.}\ }\textbf {\bibinfo {volume}
  {111}},\ \bibinfo {pages} {2020--2032} (\bibinfo {year} {2013})},\ \Eprint
  {http://arxiv.org/abs/1302.4682} {arXiv:1302.4682 [physics]} \BibitemShut
  {NoStop}%
\bibitem [{\citenamefont {Roth}, \citenamefont {Blythe},\ and\ \citenamefont
  {Schiller}(2007)}]{Roth:PRA75:023402}%
  \BibitemOpen
  \bibfield  {author} {\bibinfo {author} {\bibfnamefont {B.}~\bibnamefont
  {Roth}}, \bibinfo {author} {\bibfnamefont {P.}~\bibnamefont {Blythe}}, \ and\
  \bibinfo {author} {\bibfnamefont {S.}~\bibnamefont {Schiller}},\ }\bibfield
  {title} {\enquote {\bibinfo {title} {Motional resonance coupling in cold
  multispecies coulomb crystals},}\ }\href {\doibase
  10.1103/PhysRevA.75.023402} {\bibfield  {journal} {\bibinfo  {journal}
  {Phys.\ Rev.\ A}\ }\textbf {\bibinfo {volume} {75}},\ \bibinfo {pages}
  {023402} (\bibinfo {year} {2007})}\BibitemShut {NoStop}%
\bibitem [{\citenamefont {Reese}\ \emph {et~al.}(2004)\citenamefont {Reese},
  \citenamefont {Nguyen}, \citenamefont {Korter},\ and\ \citenamefont
  {Pratt}}]{Reese:JACS126:11387}%
  \BibitemOpen
  \bibfield  {author} {\bibinfo {author} {\bibfnamefont {J.~A.}\ \bibnamefont
  {Reese}}, \bibinfo {author} {\bibfnamefont {T.~V.}\ \bibnamefont {Nguyen}},
  \bibinfo {author} {\bibfnamefont {T.~M.}\ \bibnamefont {Korter}}, \ and\
  \bibinfo {author} {\bibfnamefont {D.~W.}\ \bibnamefont {Pratt}},\ }\bibfield
  {title} {\enquote {\bibinfo {title} {Charge redistribution on electronic
  excitation. {D}ipole moments of cis and trans 3-aminophenol in their {S}$_0$
  and {S}$_1$ electronic states},}\ }\href {\doibase 10.1021/ja0469683}
  {\bibfield  {journal} {\bibinfo  {journal} {J.\ Am.\ Chem.\ Soc.}\ }\textbf
  {\bibinfo {volume} {126}},\ \bibinfo {pages} {11387--11392} (\bibinfo {year}
  {2004})}\BibitemShut {NoStop}%
\bibitem [{\citenamefont {Frisch}\ \emph {et~al.}()\citenamefont {Frisch},
  \citenamefont {Trucks}, \citenamefont {Schlegel}, \citenamefont {Scuseria},
  \citenamefont {Robb}, \citenamefont {Cheeseman}, \citenamefont {Scalmani},
  \citenamefont {Barone}, \citenamefont {Mennucci}, \citenamefont {Petersson},
  \citenamefont {Nakatsuji}, \citenamefont {Caricato}, \citenamefont {Li},
  \citenamefont {Hratchian}, \citenamefont {Izmaylov}, \citenamefont {Bloino},
  \citenamefont {Zheng}, \citenamefont {Sonnenberg}, \citenamefont {Hada},
  \citenamefont {Ehara}, \citenamefont {Toyota}, \citenamefont {Fukuda},
  \citenamefont {Hasegawa}, \citenamefont {Ishida}, \citenamefont {Nakajima},
  \citenamefont {Honda}, \citenamefont {Kitao}, \citenamefont {Nakai},
  \citenamefont {Vreven}, \citenamefont {Montgomery}, \citenamefont {Peralta},
  \citenamefont {Ogliaro}, \citenamefont {Bearpark}, \citenamefont {Heyd},
  \citenamefont {Brothers}, \citenamefont {Kudin}, \citenamefont {Staroverov},
  \citenamefont {Kobayashi}, \citenamefont {Normand}, \citenamefont
  {Raghavachari}, \citenamefont {Rendell}, \citenamefont {Burant},
  \citenamefont {Iyengar}, \citenamefont {Tomasi}, \citenamefont {Cossi},
  \citenamefont {Rega}, \citenamefont {Millam}, \citenamefont {Klene},
  \citenamefont {Knox}, \citenamefont {Cross}, \citenamefont {Bakken},
  \citenamefont {Adamo}, \citenamefont {Jaramillo}, \citenamefont {Gomperts},
  \citenamefont {Stratmann}, \citenamefont {Yazyev}, \citenamefont {Austin},
  \citenamefont {Cammi}, \citenamefont {Pomelli}, \citenamefont {Ochterski},
  \citenamefont {Martin}, \citenamefont {Morokuma}, \citenamefont {Zakrzewski},
  \citenamefont {Voth}, \citenamefont {Salvador}, \citenamefont {Dannenberg},
  \citenamefont {Dapprich}, \citenamefont {Daniels}, \citenamefont {Farkas},
  \citenamefont {Foresman}, \citenamefont {Ortiz}, \citenamefont {Cioslowski},\
  and\ \citenamefont {Fox}}]{Gaussian:2009A02}%
  \BibitemOpen
  \bibfield  {author} {\bibinfo {author} {\bibfnamefont {M.~J.}\ \bibnamefont
  {Frisch}}, \bibinfo {author} {\bibfnamefont {G.~W.}\ \bibnamefont {Trucks}},
  \bibinfo {author} {\bibfnamefont {H.~B.}\ \bibnamefont {Schlegel}}, \bibinfo
  {author} {\bibfnamefont {G.~E.}\ \bibnamefont {Scuseria}}, \bibinfo {author}
  {\bibfnamefont {M.~A.}\ \bibnamefont {Robb}}, \bibinfo {author}
  {\bibfnamefont {J.~R.}\ \bibnamefont {Cheeseman}}, \bibinfo {author}
  {\bibfnamefont {G.}~\bibnamefont {Scalmani}}, \bibinfo {author}
  {\bibfnamefont {V.}~\bibnamefont {Barone}}, \bibinfo {author} {\bibfnamefont
  {B.}~\bibnamefont {Mennucci}}, \bibinfo {author} {\bibfnamefont {G.~A.}\
  \bibnamefont {Petersson}}, \bibinfo {author} {\bibfnamefont {H.}~\bibnamefont
  {Nakatsuji}}, \bibinfo {author} {\bibfnamefont {M.}~\bibnamefont {Caricato}},
  \bibinfo {author} {\bibfnamefont {X.}~\bibnamefont {Li}}, \bibinfo {author}
  {\bibfnamefont {H.~P.}\ \bibnamefont {Hratchian}}, \bibinfo {author}
  {\bibfnamefont {A.~F.}\ \bibnamefont {Izmaylov}}, \bibinfo {author}
  {\bibfnamefont {J.}~\bibnamefont {Bloino}}, \bibinfo {author} {\bibfnamefont
  {G.}~\bibnamefont {Zheng}}, \bibinfo {author} {\bibfnamefont {J.~L.}\
  \bibnamefont {Sonnenberg}}, \bibinfo {author} {\bibfnamefont
  {M.}~\bibnamefont {Hada}}, \bibinfo {author} {\bibfnamefont {M.}~\bibnamefont
  {Ehara}}, \bibinfo {author} {\bibfnamefont {K.}~\bibnamefont {Toyota}},
  \bibinfo {author} {\bibfnamefont {R.}~\bibnamefont {Fukuda}}, \bibinfo
  {author} {\bibfnamefont {J.}~\bibnamefont {Hasegawa}}, \bibinfo {author}
  {\bibfnamefont {M.}~\bibnamefont {Ishida}}, \bibinfo {author} {\bibfnamefont
  {T.}~\bibnamefont {Nakajima}}, \bibinfo {author} {\bibfnamefont
  {Y.}~\bibnamefont {Honda}}, \bibinfo {author} {\bibfnamefont
  {O.}~\bibnamefont {Kitao}}, \bibinfo {author} {\bibfnamefont
  {H.}~\bibnamefont {Nakai}}, \bibinfo {author} {\bibfnamefont
  {T.}~\bibnamefont {Vreven}}, \bibinfo {author} {\bibfnamefont {J.~A.}\
  \bibnamefont {Montgomery}, \bibfnamefont {{Jr.}}}, \bibinfo {author}
  {\bibfnamefont {J.~E.}\ \bibnamefont {Peralta}}, \bibinfo {author}
  {\bibfnamefont {F.}~\bibnamefont {Ogliaro}}, \bibinfo {author} {\bibfnamefont
  {M.}~\bibnamefont {Bearpark}}, \bibinfo {author} {\bibfnamefont {J.~J.}\
  \bibnamefont {Heyd}}, \bibinfo {author} {\bibfnamefont {E.}~\bibnamefont
  {Brothers}}, \bibinfo {author} {\bibfnamefont {K.~N.}\ \bibnamefont {Kudin}},
  \bibinfo {author} {\bibfnamefont {V.~N.}\ \bibnamefont {Staroverov}},
  \bibinfo {author} {\bibfnamefont {R.}~\bibnamefont {Kobayashi}}, \bibinfo
  {author} {\bibfnamefont {J.}~\bibnamefont {Normand}}, \bibinfo {author}
  {\bibfnamefont {K.}~\bibnamefont {Raghavachari}}, \bibinfo {author}
  {\bibfnamefont {A.}~\bibnamefont {Rendell}}, \bibinfo {author} {\bibfnamefont
  {J.~C.}\ \bibnamefont {Burant}}, \bibinfo {author} {\bibfnamefont {S.~S.}\
  \bibnamefont {Iyengar}}, \bibinfo {author} {\bibfnamefont {J.}~\bibnamefont
  {Tomasi}}, \bibinfo {author} {\bibfnamefont {M.}~\bibnamefont {Cossi}},
  \bibinfo {author} {\bibfnamefont {N.}~\bibnamefont {Rega}}, \bibinfo {author}
  {\bibfnamefont {J.~M.}\ \bibnamefont {Millam}}, \bibinfo {author}
  {\bibfnamefont {M.}~\bibnamefont {Klene}}, \bibinfo {author} {\bibfnamefont
  {J.~E.}\ \bibnamefont {Knox}}, \bibinfo {author} {\bibfnamefont {J.~B.}\
  \bibnamefont {Cross}}, \bibinfo {author} {\bibfnamefont {V.}~\bibnamefont
  {Bakken}}, \bibinfo {author} {\bibfnamefont {C.}~\bibnamefont {Adamo}},
  \bibinfo {author} {\bibfnamefont {J.}~\bibnamefont {Jaramillo}}, \bibinfo
  {author} {\bibfnamefont {R.}~\bibnamefont {Gomperts}}, \bibinfo {author}
  {\bibfnamefont {R.~E.}\ \bibnamefont {Stratmann}}, \bibinfo {author}
  {\bibfnamefont {O.}~\bibnamefont {Yazyev}}, \bibinfo {author} {\bibfnamefont
  {A.~J.}\ \bibnamefont {Austin}}, \bibinfo {author} {\bibfnamefont
  {R.}~\bibnamefont {Cammi}}, \bibinfo {author} {\bibfnamefont
  {C.}~\bibnamefont {Pomelli}}, \bibinfo {author} {\bibfnamefont {J.~W.}\
  \bibnamefont {Ochterski}}, \bibinfo {author} {\bibfnamefont {R.~L.}\
  \bibnamefont {Martin}}, \bibinfo {author} {\bibfnamefont {K.}~\bibnamefont
  {Morokuma}}, \bibinfo {author} {\bibfnamefont {V.~G.}\ \bibnamefont
  {Zakrzewski}}, \bibinfo {author} {\bibfnamefont {G.~A.}\ \bibnamefont
  {Voth}}, \bibinfo {author} {\bibfnamefont {P.}~\bibnamefont {Salvador}},
  \bibinfo {author} {\bibfnamefont {J.~J.}\ \bibnamefont {Dannenberg}},
  \bibinfo {author} {\bibfnamefont {S.}~\bibnamefont {Dapprich}}, \bibinfo
  {author} {\bibfnamefont {A.~D.}\ \bibnamefont {Daniels}}, \bibinfo {author}
  {\bibfnamefont {{\"O}.}~\bibnamefont {Farkas}}, \bibinfo {author}
  {\bibfnamefont {J.~B.}\ \bibnamefont {Foresman}}, \bibinfo {author}
  {\bibfnamefont {J.~V.}\ \bibnamefont {Ortiz}}, \bibinfo {author}
  {\bibfnamefont {J.}~\bibnamefont {Cioslowski}}, \ and\ \bibinfo {author}
  {\bibfnamefont {D.~J.}\ \bibnamefont {Fox}},\ }\href@noop {} {\enquote
  {\bibinfo {title} {Gaussian~09 {R}evision {A}.02},}\ }\bibinfo {note}
  {Gaussian Inc. Wallingford CT 2009}\BibitemShut {NoStop}%
\bibitem [{\citenamefont {Lynch}\ \emph {et~al.}(2000)\citenamefont {Lynch},
  \citenamefont {Fast}, \citenamefont {Harris},\ and\ \citenamefont
  {Truhlar}}]{Lynch:JPCA104:4811}%
  \BibitemOpen
  \bibfield  {author} {\bibinfo {author} {\bibfnamefont {B.~J.}\ \bibnamefont
  {Lynch}}, \bibinfo {author} {\bibfnamefont {P.~L.}\ \bibnamefont {Fast}},
  \bibinfo {author} {\bibfnamefont {M.}~\bibnamefont {Harris}}, \ and\ \bibinfo
  {author} {\bibfnamefont {D.~G.}\ \bibnamefont {Truhlar}},\ }\bibfield
  {title} {\enquote {\bibinfo {title} {Adiabatic connection for kinetics},}\
  }\href {\doibase 10.1021/jp000497z} {\bibfield  {journal} {\bibinfo
  {journal} {J.\ Phys.\ Chem.\ A}\ }\textbf {\bibinfo {volume} {104}},\
  \bibinfo {pages} {4811} (\bibinfo {year} {2000})}\BibitemShut {NoStop}%
\bibitem [{\citenamefont {Dunning}(1989)}]{Dunning:JCP90:1007}%
  \BibitemOpen
  \bibfield  {author} {\bibinfo {author} {\bibfnamefont {T.~H.}\ \bibnamefont
  {Dunning}},\ }\bibfield  {title} {\enquote {\bibinfo {title} {Gaussian basis
  sets for use in correlated molecular calculations. i. the atoms boron through
  neon and hydrogen},}\ }\href {\doibase 10.1063/1.456153} {\bibfield
  {journal} {\bibinfo  {journal} {J.\ Chem.\ Phys.}\ }\textbf {\bibinfo
  {volume} {90}},\ \bibinfo {pages} {1007} (\bibinfo {year}
  {1989})}\BibitemShut {NoStop}%
\bibitem [{\citenamefont {Peng}\ and\ \citenamefont
  {Schlegel}(1993)}]{Peng:IJC33:449}%
  \BibitemOpen
  \bibfield  {author} {\bibinfo {author} {\bibfnamefont {C.}~\bibnamefont
  {Peng}}\ and\ \bibinfo {author} {\bibfnamefont {H.~B.}\ \bibnamefont
  {Schlegel}},\ }\bibfield  {title} {\enquote {\bibinfo {title} {Combining
  synchronous transit and quasi-newton methods to find transition states},}\
  }\href {\doibase 10.1002/ijch.199300051} {\bibfield  {journal} {\bibinfo
  {journal} {Israel J.\ of Chem.}\ }\textbf {\bibinfo {volume} {33}},\ \bibinfo
  {pages} {449} (\bibinfo {year} {1993})}\BibitemShut {NoStop}%
\bibitem [{\citenamefont {Peng}\ \emph {et~al.}(1996)\citenamefont {Peng},
  \citenamefont {Ayala}, \citenamefont {Schlegel},\ and\ \citenamefont
  {Frisch}}]{Peng:JCC17:49}%
  \BibitemOpen
  \bibfield  {author} {\bibinfo {author} {\bibfnamefont {C.}~\bibnamefont
  {Peng}}, \bibinfo {author} {\bibfnamefont {P.~Y.}\ \bibnamefont {Ayala}},
  \bibinfo {author} {\bibfnamefont {H.~B.}\ \bibnamefont {Schlegel}}, \ and\
  \bibinfo {author} {\bibfnamefont {M.~J.}\ \bibnamefont {Frisch}},\ }\bibfield
   {title} {\enquote {\bibinfo {title} {{Using redundant internal coordinates
  to optimize equilibrium geometries and transition states}},}\ }\href
  {\doibase 10.1002/(SICI)1096-987X(19960115)17:1<49::AID-JCC5>3.0.CO;2-0}
  {\bibfield  {journal} {\bibinfo  {journal} {J.\ Comput.\ Chem.}\ }\textbf
  {\bibinfo {volume} {17}},\ \bibinfo {pages} {49--56} (\bibinfo {year}
  {1996})}\BibitemShut {NoStop}%
\bibitem [{\citenamefont {Clary}(1985)}]{Clary:MP54:605}%
  \BibitemOpen
  \bibfield  {author} {\bibinfo {author} {\bibfnamefont {D.~C.}\ \bibnamefont
  {Clary}},\ }\bibfield  {title} {\enquote {\bibinfo {title} {Calculations of
  rate constants for ion-molecule reactions using a combined capture and
  centrifugal sudden approximation},}\ }\href {\doibase
  10.1080/00268978500100461} {\bibfield  {journal} {\bibinfo  {journal} {Mol.\
  Phys.}\ }\textbf {\bibinfo {volume} {54}},\ \bibinfo {pages} {605} (\bibinfo
  {year} {1985})}\BibitemShut {NoStop}%
\bibitem [{\citenamefont {Stoecklin}, \citenamefont {Clary},\ and\
  \citenamefont {Palma}(1992)}]{Stoecklin:JCSFT88:901}%
  \BibitemOpen
  \bibfield  {author} {\bibinfo {author} {\bibfnamefont {T.}~\bibnamefont
  {Stoecklin}}, \bibinfo {author} {\bibfnamefont {D.~C.}\ \bibnamefont
  {Clary}}, \ and\ \bibinfo {author} {\bibfnamefont {A.}~\bibnamefont
  {Palma}},\ }\bibfield  {title} {\enquote {\bibinfo {title} {Rate constant
  calculations for ion-symmetric top and ion-asymmetric top reactions},}\
  }\href {\doibase 10.1039/FT9928800901} {\bibfield  {journal} {\bibinfo
  {journal} {J.\ Chem.\ Soc.\ -- Faraday Trans.}\ }\textbf {\bibinfo {volume}
  {88}},\ \bibinfo {pages} {901} (\bibinfo {year} {1992})}\BibitemShut
  {NoStop}%
\bibitem [{Note1()}]{Note1}%
  \BibitemOpen
  \bibinfo {note} {Here, ``permanent dipole'' refers to the dipole moment of
  \protect \ensuremath {\protect \text {3AP}}\protect \xspace in its molecular
  frame~[\protect \rev@citealpnum {Klemperer:JPC97:2413}].}\BibitemShut {Stop}%
\bibitem [{\citenamefont {Filsinger}\ \emph
  {et~al.}(2008{\natexlab{b}})\citenamefont {Filsinger}, \citenamefont
  {Wohlfart}, \citenamefont {Schnell}, \citenamefont {Grabow},\ and\
  \citenamefont {K\"upper}}]{Filsinger:PCCP10:666}%
  \BibitemOpen
  \bibfield  {author} {\bibinfo {author} {\bibfnamefont {F.}~\bibnamefont
  {Filsinger}}, \bibinfo {author} {\bibfnamefont {K.}~\bibnamefont {Wohlfart}},
  \bibinfo {author} {\bibfnamefont {M.}~\bibnamefont {Schnell}}, \bibinfo
  {author} {\bibfnamefont {J.-U.}\ \bibnamefont {Grabow}}, \ and\ \bibinfo
  {author} {\bibfnamefont {J.}~\bibnamefont {K\"upper}},\ }\bibfield  {title}
  {\enquote {\bibinfo {title} {Precise dipole moments and quadrupole coupling
  constants of the cis and trans conformers of 3-aminophenol: Determination of
  the absolute conformation},}\ }\href {\doibase 10.1039/b711888k} {\bibfield
  {journal} {\bibinfo  {journal} {Phys.\ Chem.\ Chem.\ Phys.}\ }\textbf
  {\bibinfo {volume} {10}},\ \bibinfo {pages} {666--673} (\bibinfo {year}
  {2008}{\natexlab{b}})},\ \Eprint {http://arxiv.org/abs/0708.0282}
  {arXiv:0708.0282 [physics]} \BibitemShut {NoStop}%
\bibitem [{\citenamefont {Matthey}\ \emph {et~al.}(2004)\citenamefont
  {Matthey}, \citenamefont {Cickovski}, \citenamefont {Hampton}, \citenamefont
  {Ko}, \citenamefont {Ma}, \citenamefont {Nyerges}, \citenamefont {Raeder},
  \citenamefont {Slabach},\ and\ \citenamefont
  {Izaguirre}}]{Matthey:ATMS30:237}%
  \BibitemOpen
  \bibfield  {author} {\bibinfo {author} {\bibfnamefont {T.}~\bibnamefont
  {Matthey}}, \bibinfo {author} {\bibfnamefont {T.}~\bibnamefont {Cickovski}},
  \bibinfo {author} {\bibfnamefont {S.}~\bibnamefont {Hampton}}, \bibinfo
  {author} {\bibfnamefont {A.}~\bibnamefont {Ko}}, \bibinfo {author}
  {\bibfnamefont {Q.}~\bibnamefont {Ma}}, \bibinfo {author} {\bibfnamefont
  {M.}~\bibnamefont {Nyerges}}, \bibinfo {author} {\bibfnamefont
  {T.}~\bibnamefont {Raeder}}, \bibinfo {author} {\bibfnamefont
  {T.}~\bibnamefont {Slabach}}, \ and\ \bibinfo {author} {\bibfnamefont
  {J.~A.}\ \bibnamefont {Izaguirre}},\ }\bibfield  {title} {\enquote {\bibinfo
  {title} {Protomol, an object-oriented framework for prototyping novel
  algorithms for molecular dynamics},}\ }\href {\doibase
  10.1145/1024074.1024075} {\bibfield  {journal} {\bibinfo  {journal} {ACM
  Trans.\ Math.\ Softw.}\ }\textbf {\bibinfo {volume} {30}},\ \bibinfo {pages}
  {237--265} (\bibinfo {year} {2004})}\BibitemShut {NoStop}%
\bibitem [{\citenamefont {Bell}\ \emph {et~al.}(2009)\citenamefont {Bell},
  \citenamefont {Gingell}, \citenamefont {Oldham}, \citenamefont {Softley},\
  and\ \citenamefont {Willitsch}}]{Bell:FD142:73}%
  \BibitemOpen
  \bibfield  {author} {\bibinfo {author} {\bibfnamefont {M.~T.}\ \bibnamefont
  {Bell}}, \bibinfo {author} {\bibfnamefont {A.~D.}\ \bibnamefont {Gingell}},
  \bibinfo {author} {\bibfnamefont {J.}~\bibnamefont {Oldham}}, \bibinfo
  {author} {\bibfnamefont {T.~P.}\ \bibnamefont {Softley}}, \ and\ \bibinfo
  {author} {\bibfnamefont {S.}~\bibnamefont {Willitsch}},\ }\bibfield  {title}
  {\enquote {\bibinfo {title} {Ion-molecule chemistry at very low temperatures:
  cold chemical reactions between {C}oulomb-crystallized ions and
  velocity-selected neutral molecules},}\ }\href {\doibase 10.1039/B818733A}
  {\bibfield  {journal} {\bibinfo  {journal} {Faraday Disc.}\ }\textbf
  {\bibinfo {volume} {142}},\ \bibinfo {pages} {73} (\bibinfo {year}
  {2009})}\BibitemShut {NoStop}%
\bibitem [{\citenamefont {K\"upper}, \citenamefont {Filsinger},\ and\
  \citenamefont {Meijer}(2009)}]{Kuepper:FD142:155}%
  \BibitemOpen
  \bibfield  {author} {\bibinfo {author} {\bibfnamefont {J.}~\bibnamefont
  {K\"upper}}, \bibinfo {author} {\bibfnamefont {F.}~\bibnamefont {Filsinger}},
  \ and\ \bibinfo {author} {\bibfnamefont {G.}~\bibnamefont {Meijer}},\
  }\bibfield  {title} {\enquote {\bibinfo {title} {Manipulating the motion of
  large molecules},}\ }\href {\doibase 10.1039/b820045a} {\bibfield  {journal}
  {\bibinfo  {journal} {Faraday Disc.}\ }\textbf {\bibinfo {volume} {142}},\
  \bibinfo {pages} {155--173} (\bibinfo {year} {2009})},\ \Eprint
  {http://arxiv.org/abs/0906.4355} {arXiv:0906.4355 [physics]} \BibitemShut
  {NoStop}%
\bibitem [{\citenamefont {Chang}\ \emph {et~al.}(2014)\citenamefont {Chang},
  \citenamefont {Filsinger}, \citenamefont {Sartakov},\ and\ \citenamefont
  {K{\"u}pper}}]{Chang:CPC185:339}%
  \BibitemOpen
  \bibfield  {author} {\bibinfo {author} {\bibfnamefont {Y.-P.}\ \bibnamefont
  {Chang}}, \bibinfo {author} {\bibfnamefont {F.}~\bibnamefont {Filsinger}},
  \bibinfo {author} {\bibfnamefont {B.}~\bibnamefont {Sartakov}}, \ and\
  \bibinfo {author} {\bibfnamefont {J.}~\bibnamefont {K{\"u}pper}},\ }\bibfield
   {title} {\enquote {\bibinfo {title} {\textsc{CMIstark}\xspace : Python
  package for the stark-effect calculation and symmetry classification of
  linear, symmetric and asymmetric top wavefunctions in dc electric fields},}\
  }\href {\doibase http://dx.doi.org/10.1016/j.cpc.2013.09.001} {\bibfield
  {journal} {\bibinfo  {journal} {Comp.\ Phys.\ Comm.}\ }\textbf {\bibinfo
  {volume} {185}},\ \bibinfo {pages} {339--49} (\bibinfo {year} {2014})},\
  \Eprint {http://arxiv.org/abs/1308.4076} {arXiv:1308.4076 [physics]}
  \BibitemShut {NoStop}%
\bibitem [{\citenamefont {Plane}\ \emph {et~al.}(2006)\citenamefont {Plane},
  \citenamefont {Vondrak}, \citenamefont {Broadley}, \citenamefont {Cosic},
  \citenamefont {Ermoline},\ and\ \citenamefont
  {Fontijn}}]{Plane:JPCA110:7874}%
  \BibitemOpen
  \bibfield  {author} {\bibinfo {author} {\bibfnamefont {J.~M.~C.}\
  \bibnamefont {Plane}}, \bibinfo {author} {\bibfnamefont {T.}~\bibnamefont
  {Vondrak}}, \bibinfo {author} {\bibfnamefont {S.}~\bibnamefont {Broadley}},
  \bibinfo {author} {\bibfnamefont {B.}~\bibnamefont {Cosic}}, \bibinfo
  {author} {\bibfnamefont {A.}~\bibnamefont {Ermoline}}, \ and\ \bibinfo
  {author} {\bibfnamefont {A.}~\bibnamefont {Fontijn}},\ }\bibfield  {title}
  {\enquote {\bibinfo {title} {Kinetic study of the reaction {Ca$^+$} +
  {N$_2$O} from 188 to 1207~k},}\ }\href {\doibase 10.1021/jp061664j}
  {\bibfield  {journal} {\bibinfo  {journal} {J.\ Phys.\ Chem.\ A}\ }\textbf
  {\bibinfo {volume} {110}},\ \bibinfo {pages} {7874} (\bibinfo {year}
  {2006})}\BibitemShut {NoStop}%
\bibitem [{\citenamefont {Chickos}\ and\ \citenamefont
  {Acree}(2002)}]{Chickos:JPCRD31:537}%
  \BibitemOpen
  \bibfield  {author} {\bibinfo {author} {\bibfnamefont {J.~S.}\ \bibnamefont
  {Chickos}}\ and\ \bibinfo {author} {\bibfnamefont {W.~E.}\ \bibnamefont
  {Acree}},\ }\bibfield  {title} {\enquote {\bibinfo {title} {Enthalpies of
  sublimation of organic and organometallic compounds. 1910-2001},}\ }\href
  {\doibase http://dx.doi.org/10.1063/1.1475333} {\bibfield  {journal}
  {\bibinfo  {journal} {J.\ Phys.\ Chem.\ Ref.\ Data}\ }\textbf {\bibinfo
  {volume} {31}},\ \bibinfo {pages} {537} (\bibinfo {year} {2002})}\BibitemShut
  {NoStop}%
\bibitem [{Note2()}]{Note2}%
  \BibitemOpen
  \bibinfo {note} {For \protect \ensuremath {\protect \text {3AP}}\protect
  \xspace , the number density in each gas pulse $n_\protect \text {pulse}$ is
  equal to $2.56(41)\times 10^{8}$~cm$^{-3}$. $n_\protect \text {avg}$ equals
  to a product of the gas pulse duration (50~$\mu $s), the repetition rate
  (600~Hz), and $n_\protect \text {pulse}$.}\BibitemShut {Stop}%
\bibitem [{\citenamefont {Rowe}\ and\ \citenamefont
  {Marquette}(1987)}]{Rowe:IJMSIP80:239}%
  \BibitemOpen
  \bibfield  {author} {\bibinfo {author} {\bibfnamefont {B.~R.}\ \bibnamefont
  {Rowe}}\ and\ \bibinfo {author} {\bibfnamefont {J.~B.}\ \bibnamefont
  {Marquette}},\ }\bibfield  {title} {\enquote {\bibinfo {title} {{CRESU}
  studies of ion/moelcule reactions},}\ }\href {\doibase
  10.1016/0168-1176(87)87033-7} {\bibfield  {journal} {\bibinfo  {journal}
  {Int.\ J.\ Mass Spectrom.\ Ion Proc.}\ }\textbf {\bibinfo {volume} {80}},\
  \bibinfo {pages} {239} (\bibinfo {year} {1987})}\BibitemShut {NoStop}%
\bibitem [{\citenamefont {Clary}(1990)}]{Clary:ARPC41:61}%
  \BibitemOpen
  \bibfield  {author} {\bibinfo {author} {\bibfnamefont {D.}~\bibnamefont
  {Clary}},\ }\bibfield  {title} {\enquote {\bibinfo {title} {{Fast Chemical
  Reactions: Theory Challenges Experiment}},}\ }\href {\doibase
  10.1146/annurev.physchem.41.1.61} {\bibfield  {journal} {\bibinfo  {journal}
  {Annu.\ Rev.\ Phys.\ Chem.}\ }\textbf {\bibinfo {volume} {41}},\ \bibinfo
  {pages} {61--90} (\bibinfo {year} {1990})}\BibitemShut {NoStop}%
\bibitem [{\citenamefont {Sabbah}\ \emph {et~al.}(2007)\citenamefont {Sabbah},
  \citenamefont {Biennier}, \citenamefont {Sims}, \citenamefont {Georgievskii},
  \citenamefont {Klippenstein},\ and\ \citenamefont
  {Smith}}]{Sabbah:Science317:102}%
  \BibitemOpen
  \bibfield  {author} {\bibinfo {author} {\bibfnamefont {H.}~\bibnamefont
  {Sabbah}}, \bibinfo {author} {\bibfnamefont {L.}~\bibnamefont {Biennier}},
  \bibinfo {author} {\bibfnamefont {I.~R.}\ \bibnamefont {Sims}}, \bibinfo
  {author} {\bibfnamefont {Y.}~\bibnamefont {Georgievskii}}, \bibinfo {author}
  {\bibfnamefont {S.~J.}\ \bibnamefont {Klippenstein}}, \ and\ \bibinfo
  {author} {\bibfnamefont {I.~W.~M.}\ \bibnamefont {Smith}},\ }\bibfield
  {title} {\enquote {\bibinfo {title} {{Understanding reactivity at very low
  temperatures: the reactions of oxygen atoms with alkenes.}}}\ }\href
  {\doibase 10.1126/science.1142373} {\bibfield  {journal} {\bibinfo  {journal}
  {Science}\ }\textbf {\bibinfo {volume} {317}},\ \bibinfo {pages} {102--5}
  (\bibinfo {year} {2007})}\BibitemShut {NoStop}%
\bibitem [{\citenamefont {Gingell}\ \emph {et~al.}(2010)\citenamefont
  {Gingell}, \citenamefont {Bell}, \citenamefont {Oldham}, \citenamefont
  {Softley},\ and\ \citenamefont {Harvey}}]{Gingell:JCP133:194302}%
  \BibitemOpen
  \bibfield  {author} {\bibinfo {author} {\bibfnamefont {A.~D.}\ \bibnamefont
  {Gingell}}, \bibinfo {author} {\bibfnamefont {M.~T.}\ \bibnamefont {Bell}},
  \bibinfo {author} {\bibfnamefont {J.~M.}\ \bibnamefont {Oldham}}, \bibinfo
  {author} {\bibfnamefont {T.~P.}\ \bibnamefont {Softley}}, \ and\ \bibinfo
  {author} {\bibfnamefont {J.~N.}\ \bibnamefont {Harvey}},\ }\bibfield  {title}
  {\enquote {\bibinfo {title} {Cold chemistry with electronically excited
  {Ca$^+$} {C}oulomb crystals.}}\ }\href {\doibase 10.1063/1.3505142}
  {\bibfield  {journal} {\bibinfo  {journal} {J.\ Chem.\ Phys.}\ }\textbf
  {\bibinfo {volume} {133}},\ \bibinfo {pages} {194302} (\bibinfo {year}
  {2010})}\BibitemShut {NoStop}%
\bibitem [{\citenamefont {Smith}\ \emph {et~al.}(2006)\citenamefont {Smith},
  \citenamefont {Sage}, \citenamefont {Donahue}, \citenamefont {Herbst},\ and\
  \citenamefont {Quan}}]{Smith:FD133:137}%
  \BibitemOpen
  \bibfield  {author} {\bibinfo {author} {\bibfnamefont {I.~W.~M.}\
  \bibnamefont {Smith}}, \bibinfo {author} {\bibfnamefont {A.~M.}\ \bibnamefont
  {Sage}}, \bibinfo {author} {\bibfnamefont {N.~M.}\ \bibnamefont {Donahue}},
  \bibinfo {author} {\bibfnamefont {E.}~\bibnamefont {Herbst}}, \ and\ \bibinfo
  {author} {\bibfnamefont {D.}~\bibnamefont {Quan}},\ }\bibfield  {title}
  {\enquote {\bibinfo {title} {{The temperature-dependence of rapid low
  temperature reactions: experiment, understanding and prediction}},}\ }\href
  {\doibase 10.1039/b600721j} {\bibfield  {journal} {\bibinfo  {journal}
  {Faraday Disc.}\ }\textbf {\bibinfo {volume} {133}},\ \bibinfo {pages} {137}
  (\bibinfo {year} {2006})}\BibitemShut {NoStop}%
\bibitem [{\citenamefont {Nielsen}\ \emph {et~al.}(2011)\citenamefont
  {Nielsen}, \citenamefont {Simesen}, \citenamefont {Bisgaard}, \citenamefont
  {Stapelfeldt}, \citenamefont {Filsinger}, \citenamefont {Friedrich},
  \citenamefont {Meijer},\ and\ \citenamefont
  {K\"upper}}]{Nielsen:PCCP13:18971}%
  \BibitemOpen
  \bibfield  {author} {\bibinfo {author} {\bibfnamefont {J.~H.}\ \bibnamefont
  {Nielsen}}, \bibinfo {author} {\bibfnamefont {P.}~\bibnamefont {Simesen}},
  \bibinfo {author} {\bibfnamefont {C.~Z.}\ \bibnamefont {Bisgaard}}, \bibinfo
  {author} {\bibfnamefont {H.}~\bibnamefont {Stapelfeldt}}, \bibinfo {author}
  {\bibfnamefont {F.}~\bibnamefont {Filsinger}}, \bibinfo {author}
  {\bibfnamefont {B.}~\bibnamefont {Friedrich}}, \bibinfo {author}
  {\bibfnamefont {G.}~\bibnamefont {Meijer}}, \ and\ \bibinfo {author}
  {\bibfnamefont {J.}~\bibnamefont {K\"upper}},\ }\bibfield  {title} {\enquote
  {\bibinfo {title} {Stark-selected beam of ground-state {OCS} molecules
  characterized by revivals of impulsive alignment},}\ }\href {\doibase
  10.1039/c1cp21143a} {\bibfield  {journal} {\bibinfo  {journal} {Phys.\ Chem.\
  Chem.\ Phys.}\ }\textbf {\bibinfo {volume} {13}},\ \bibinfo {pages}
  {18971--18975} (\bibinfo {year} {2011})},\ \Eprint
  {http://arxiv.org/abs/1105.2413} {arXiv:1105.2413 [physics]} \BibitemShut
  {NoStop}%
\bibitem [{\citenamefont {Putzke}\ \emph {et~al.}(2011)\citenamefont {Putzke},
  \citenamefont {Filsinger}, \citenamefont {Haak}, \citenamefont {K\"upper},\
  and\ \citenamefont {Meijer}}]{Putzke:PCCP13:18962}%
  \BibitemOpen
  \bibfield  {author} {\bibinfo {author} {\bibfnamefont {S.}~\bibnamefont
  {Putzke}}, \bibinfo {author} {\bibfnamefont {F.}~\bibnamefont {Filsinger}},
  \bibinfo {author} {\bibfnamefont {H.}~\bibnamefont {Haak}}, \bibinfo {author}
  {\bibfnamefont {J.}~\bibnamefont {K\"upper}}, \ and\ \bibinfo {author}
  {\bibfnamefont {G.}~\bibnamefont {Meijer}},\ }\bibfield  {title} {\enquote
  {\bibinfo {title} {Rotational-state-specific guiding of large molecules},}\
  }\href {\doibase 10.1039/C1CP20721K} {\bibfield  {journal} {\bibinfo
  {journal} {Phys.\ Chem.\ Chem.\ Phys.}\ }\textbf {\bibinfo {volume} {13}},\
  \bibinfo {pages} {18962} (\bibinfo {year} {2011})},\ \Eprint
  {http://arxiv.org/abs/1103.5080} {arXiv:1103.5080 [physics]} \BibitemShut
  {NoStop}%
\bibitem [{\citenamefont {Trippel}\ \emph {et~al.}(2012)\citenamefont
  {Trippel}, \citenamefont {Chang}, \citenamefont {Stern}, \citenamefont
  {Mullins}, \citenamefont {Holmegaard},\ and\ \citenamefont
  {K{\"u}pper}}]{Trippel:PRA86:033202}%
  \BibitemOpen
  \bibfield  {author} {\bibinfo {author} {\bibfnamefont {S.}~\bibnamefont
  {Trippel}}, \bibinfo {author} {\bibfnamefont {Y.-P.}\ \bibnamefont {Chang}},
  \bibinfo {author} {\bibfnamefont {S.}~\bibnamefont {Stern}}, \bibinfo
  {author} {\bibfnamefont {T.}~\bibnamefont {Mullins}}, \bibinfo {author}
  {\bibfnamefont {L.}~\bibnamefont {Holmegaard}}, \ and\ \bibinfo {author}
  {\bibfnamefont {J.}~\bibnamefont {K{\"u}pper}},\ }\bibfield  {title}
  {\enquote {\bibinfo {title} {Spatial separation of state- and size-selected
  neutral clusters},}\ }\href {\doibase 10.1103/PhysRevA.86.033202} {\bibfield
  {journal} {\bibinfo  {journal} {Phys.\ Rev.\ A}\ }\textbf {\bibinfo {volume}
  {86}},\ \bibinfo {pages} {033202} (\bibinfo {year} {2012})},\ \Eprint
  {http://arxiv.org/abs/1208.4935} {arXiv:1208.4935 [physics]} \BibitemShut
  {NoStop}%
\bibitem [{\citenamefont {van~de Meerakker}\ \emph {et~al.}(2012)\citenamefont
  {van~de Meerakker}, \citenamefont {Bethlem}, \citenamefont {Vanhaecke},\ and\
  \citenamefont {Meijer}}]{Meerakker:CR112:2012}%
  \BibitemOpen
  \bibfield  {author} {\bibinfo {author} {\bibfnamefont {S.~Y.~T.}\
  \bibnamefont {van~de Meerakker}}, \bibinfo {author} {\bibfnamefont {H.~L.}\
  \bibnamefont {Bethlem}}, \bibinfo {author} {\bibfnamefont {N.}~\bibnamefont
  {Vanhaecke}}, \ and\ \bibinfo {author} {\bibfnamefont {G.}~\bibnamefont
  {Meijer}},\ }\bibfield  {title} {\enquote {\bibinfo {title} {Manipulation and
  control of molecular beams},}\ }\href {\doibase 10.1021/cr200349r} {\bibfield
   {journal} {\bibinfo  {journal} {Chem.\ Rev.}\ }\textbf {\bibinfo {volume}
  {112}},\ \bibinfo {pages} {4828--4878} (\bibinfo {year} {2012})}\BibitemShut
  {NoStop}%
\bibitem [{\citenamefont {Tong}, \citenamefont {Winney},\ and\ \citenamefont
  {Willitsch}(2010)}]{Tong:PRL105:143001}%
  \BibitemOpen
  \bibfield  {author} {\bibinfo {author} {\bibfnamefont {X.}~\bibnamefont
  {Tong}}, \bibinfo {author} {\bibfnamefont {A.~H.}\ \bibnamefont {Winney}}, \
  and\ \bibinfo {author} {\bibfnamefont {S.}~\bibnamefont {Willitsch}},\
  }\bibfield  {title} {\enquote {\bibinfo {title} {Sympathetic cooling of
  molecular ions in selected rotational and vibrational states produced by
  threshold photoionization},}\ }\href {\doibase
  10.1103/PhysRevLett.105.143001} {\bibfield  {journal} {\bibinfo  {journal}
  {Phys.\ Rev.\ Lett.}\ }\textbf {\bibinfo {volume} {105}},\ \bibinfo {pages}
  {143001} (\bibinfo {year} {2010})},\ \Eprint {http://arxiv.org/abs/1006.5642}
  {arXiv:1006.5642 [physics]} \BibitemShut {NoStop}%
\bibitem [{\citenamefont {Tong}, \citenamefont {Wild},\ and\ \citenamefont
  {Willitsch}(2011)}]{Tong:PRA83:023415}%
  \BibitemOpen
  \bibfield  {author} {\bibinfo {author} {\bibfnamefont {X.}~\bibnamefont
  {Tong}}, \bibinfo {author} {\bibfnamefont {D.}~\bibnamefont {Wild}}, \ and\
  \bibinfo {author} {\bibfnamefont {S.}~\bibnamefont {Willitsch}},\ }\bibfield
  {title} {\enquote {\bibinfo {title} {Collisional and radiative effects in the
  state-selective preparation of translationally cold molecular ions in ion
  traps},}\ }\href {\doibase 10.1103/PhysRevA.83.023415} {\bibfield  {journal}
  {\bibinfo  {journal} {Phys.\ Rev.\ A}\ }\textbf {\bibinfo {volume} {83}},\
  \bibinfo {pages} {023415} (\bibinfo {year} {2011})}\BibitemShut {NoStop}%
\bibitem [{\citenamefont {Klemperer}\ \emph {et~al.}(1993)\citenamefont
  {Klemperer}, \citenamefont {Lehmann}, \citenamefont {Watson},\ and\
  \citenamefont {Wofsy}}]{Klemperer:JPC97:2413}%
  \BibitemOpen
  \bibfield  {author} {\bibinfo {author} {\bibfnamefont {W.}~\bibnamefont
  {Klemperer}}, \bibinfo {author} {\bibfnamefont {K.~K.}\ \bibnamefont
  {Lehmann}}, \bibinfo {author} {\bibfnamefont {J.~K.~G.}\ \bibnamefont
  {Watson}}, \ and\ \bibinfo {author} {\bibfnamefont {S.~C.}\ \bibnamefont
  {Wofsy}},\ }\bibfield  {title} {\enquote {\bibinfo {title} {Can molecules
  have permanent electric dipole moments?}}\ }\href {\doibase
  10.1021/j100112a049} {\bibfield  {journal} {\bibinfo  {journal} {J. Phys.
  Chem.}\ }\textbf {\bibinfo {volume} {97}},\ \bibinfo {pages} {2413--2416}
  (\bibinfo {year} {1993})}\BibitemShut {NoStop}%
\end{thebibliography}

%

\end{document}